\documentclass[reprint,superscriptaddress,amsmath,amssymb,aps,prl]{revtex4-2}
\usepackage{graphicx}
\usepackage{dcolumn}
\usepackage{bm}
\usepackage{color}
\usepackage{units}
\usepackage{wasysym}
\usepackage{MnSymbol}
\usepackage{comment}
\usepackage{times}
\usepackage{upgreek}
\usepackage{booktabs}
\usepackage{float}
\usepackage [english]{babel}
\newcommand{\fakesection}[1]{
  \par\refstepcounter{section}
  \sectionmark{#1}
  \addcontentsline{toc}{section}{\protect\numberline{\thesection}#1}
}
\usepackage [autostyle, english = american]{csquotes}
\MakeOuterQuote{"}
\usepackage[mathlines]{lineno}
\usepackage[unicode=true,pdfusetitle,bookmarks=false,colorlinks=true,citecolor=blue,urlcolor=blue,linkcolor=blue]{hyperref}
\begin{document}

\title{Hexatic Order Coupled with Thermal Noise Produces Bubbles in Two-Dimensional Active Matter}

\author{Luke Langford}
\affiliation{Department of Materials Science and Engineering, University of California, Berkeley, California 94720, USA}

\author{Ahmad K. Omar}
\email{aomar@berkeley.edu}
\affiliation{Department of Materials Science and Engineering, University of California, Berkeley, California 94720, USA}
\affiliation{Materials Sciences Division, Lawrence Berkeley National Laboratory, Berkeley, California 94720, USA}

\begin{abstract}
The phase separation of purely-repulsive particles induced by self-propulsion is among the most well-studied non-equilibrium phase transitions.
However, some notable features of this transition remain open questions, including the origin of bubbles within the dense phase in two dimensions. 
Various explanations have been proposed, ranging from a reversal of the Ostwald ripening process to topological defects at the borders of hexatic domains. 
We present particle-based simulations that disentangle the effect of hexatic domains on the bubble size and number distribution through the introduction of polydispersity. 
While hexatic order is found to be necessary for bubble formation, we also identify thermal translational noise is required for bubble generation. 
Intriguingly, the magnitude of the thermal noise needed for bubble formation can be remarkably small in comparison with the particle activity but cannot be identically zero. 
The cooperative motion evidenced within the dense phase of the thermal hexatic domains may may be necessary for bubble production. 

\end{abstract}

\maketitle

\textit{Introduction.--} 
The ubiquity of phase separation observed in environments driven far from equilibrium, in both synthetic and biological~\cite{Brangwynne2009GermlineDissolution/Condensation,Palacci2013LivingSurfers,Han2017EffectiveMixture,Liu2019Self-DrivenFormation,Soni2019TheFluid,Geyer2019FreezingLiquids,Zhang2021ActiveDensity} contexts, has inspired theoretical efforts to understand purely non-equilibrium driving forces for phase transitions.
One identified driving force is that of \textit{self-propulsion}: motile particles with purely repulsive interactions can phase separate into dense and dilute regions at sufficient density and propulsive speed, a phenomenon known as motility induced phase separation (MIPS)~\cite{Cates2015}.
Considerable recent work towards understanding MIPS has come from both modified Cahn-Hilliard frameworks~\cite{Wittkowski2014ScalarSeparation,Tjhung2018} as well as particle-based simulations~\cite{Fily2012,Redner2013StructureFluid,Stenhammar2014PhaseDimensionality}.
These combined approaches have made significant progress towards understanding the non-equilibrium behavior of phase coexistence~\cite{Solon2018GeneralizedEnsembles,Solon2018GeneralizedMatter,Digregorio2018FullSeparation,Omar2021,Omar2023b}, capillary fluctuations~\cite{Patch2018,Fausti2021CapillarySeparation,Langford2024TheoryPhases}, nucleation~\cite{Redner2016,Cates2023ClassicalSeparation,Langford2025TheMatter}, and crystallization~\cite{Evans2026TheorySpheres}.

Despite this progress, the origin of the bubbles decorating the dense phase of phase separated active particles in two dimensions~\cite{Stenhammar2014PhaseDimensionality} remains an open question.
The phenomenology of these bubbles resembles that predicted by Tjhung \textit{et al}.~\cite{Tjhung2018} using a modified Cahn-Hilliard framework Active Model B+ (AMB+).
In their minimal scalar field theory, they identified a range of parameters where the surface tension governing nucleation becomes negative. 
Under these conditions, AMB+ predicts a reversal of the Ostwald process resulting in a dense phase in which bubbles spontaneously form, coalesce, migrate to the boundaries of the dense phase, and merge with the surrounding vapor (liquid droplets dispersed in a vapor phase is also possible).
The qualitative similarities between this prediction and the bubbles found in two-dimensional particle simulations~\cite{Redner2013StructureFluid,Stenhammar2014PhaseDimensionality,Bialke2015,Caporusso2020Motility-InducedSystem,Nakano2024UniversalParticles} have led some to argue that the Ostwald process is reversed in active particles~\cite{Cates2025ActivePhysics}, although this has yet to be directly established.

\begin{figure*}[t]
	\centering
	\includegraphics[width=0.95\textwidth]{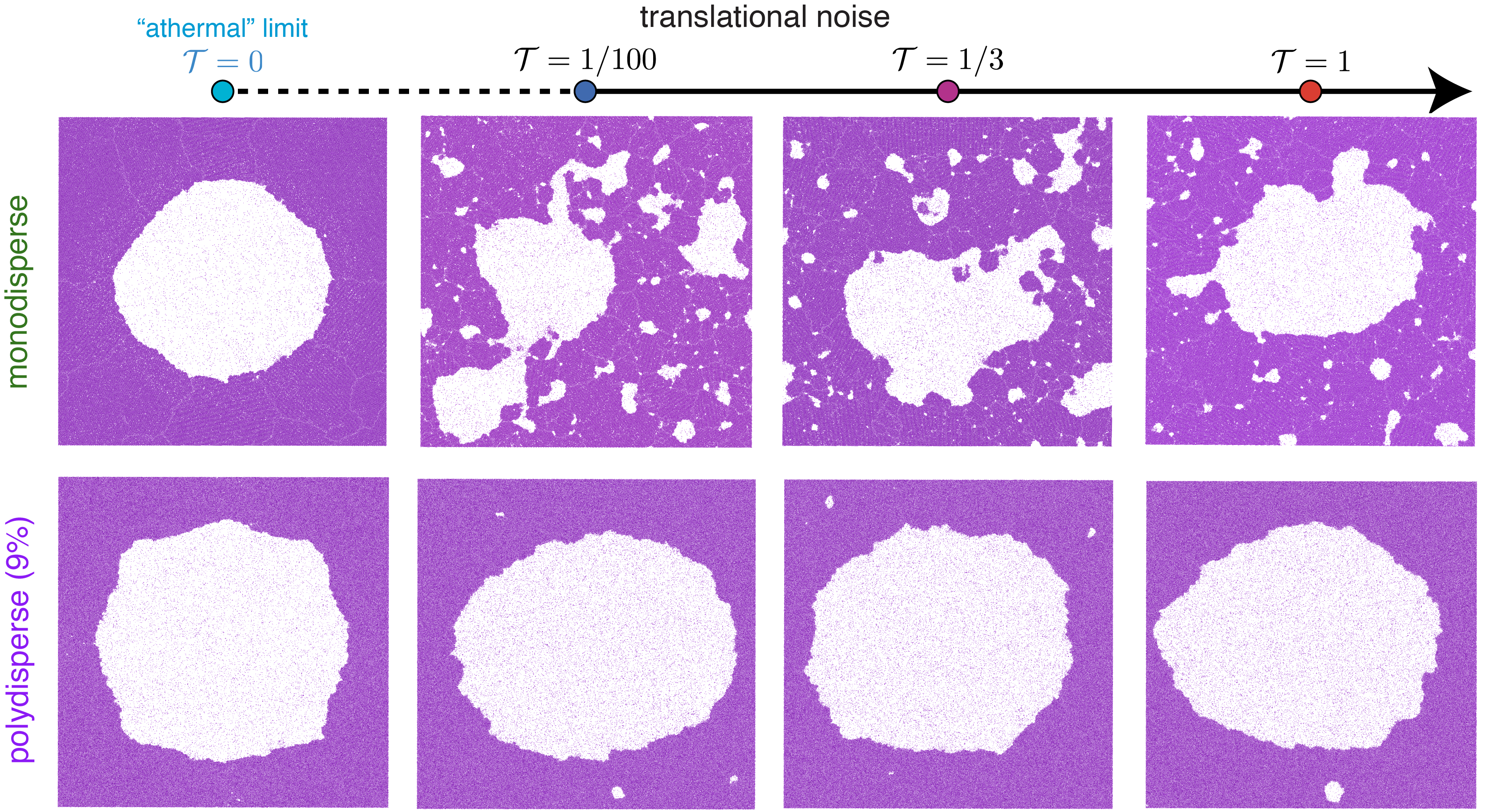}
	\caption{\protect\small{{Representative snapshots of simulations at various values of thermalization for both monodisperse and polydisperse cases, rendered using Ovito~\cite{Stukowski2010VisualizationTool}. Bubbles within the liquid phase were uncommon in all polydisperse and all athermal simulations.}}}
	\label{fig:figure1}
\end{figure*}

Caporusso \textit{et al.}~\cite{Caporusso2020Motility-InducedSystem} measured the distribution of bubble sizes in active particle simulations to be power-law distributed rather than peaked about a stable critical radius as one might naively assume a reversed Ostwald process implies.
Further, they find that bubbles are present almost exclusively at the borders of hexatic domains, implying that the bubbles may originate from topological defects rather than exotic coarsening dynamics.
However, Fausti \textit{et al.}~\cite{Fausti2024StatisticalFluids} demonstrated through numerical integration of AMB+ that the distribution of bubble sizes can be anything between narrowly peaked or power-law decaying depending on the relative strengths of reversed Ostwald ripening, coalescence dynamics, and the nucleation rate.
The numerical work of Shi \textit{et al.}~\cite{Shi2020Self-OrganizedParticles} investigated the system-size dependence of the distribution of bubbles in two-dimensional active particles, finding evidence that the bubbly fluid displays self-organized criticality (SOC). 
They also demonstrated that a reduced numerical model implementing reversed Ostwald ripening also displays SOC, further strengthening the argument that the Ostwald process is reversed in active particles.
Recently, Yan \textit{et al.} measured bubble \textit{dynamics}~\cite{Yan2025StochasticMatter} from particle simulations in addition to the size distribution.
In particular, they fit a stochastic differential equation for the size dynamics of bubbles, finding that the drift term is area-independent and strictly negative, driving bubble collapse. 
While bubble dissolution in AMB+ under reversed Ostwald ripening is driven by bubble migration to the interface,  Yan \textit{et al.} found that $97\%$ of bubbles in particle simulations spontaneously dissolve into the dense phase (even after accounting for coalescence) rather than escaping into the vapor.
Finally, bubbles have not been observed in large-scale simulations of ABPs in three dimensions~\cite{Stenhammar2014PhaseDimensionality,Omar2021}. 
Our recent theoretical work on active nucleation of ABPs in three dimensions is consistent with these numerical findings~\cite{Langford2025TheMatter}. 

Clearly, the origin of bubble formation in microscopic models of two-dimensional active particles remains an open question.
In this Letter, we present additional simulation data that further elucidates the dependencies of bubble formation. 
While previous simulation studies have focused on the dependence of the bubbles on activity, overall density, and system size, we focus on two other dependencies: polydispersity of particle size and thermalization (i.e.,~\textit{translational} noise).
Increasing polydispersity allows for the systematic elimination of hexatic order which was implicated in bubble formation in Ref.~\cite{Caporusso2020Motility-InducedSystem}.
The reduction in hexatic order is indeed found to eliminate appreciable bubble formation. 
Intriguingly the presence of hexatic domains appears to be a necessary but insufficient condition for bubble formation. 
Translational thermal noise --- while thought to play only a quantitative role in the phenomenology of MIPS --- is also necessary for appreciable bubble formation. 
Shockingly, we find that even a vanishingly small magnitude of thermalization is sufficient for the liquid to be populated with bubbles -- only when translational noise is precisely zero do bubbles become scarce.
These findings provide an increasingly complete portrait regarding the nature of bubbles in two-dimensional active matter.

\textit{Model System and Numerical Details.--}
We conduct simulations of two-dimensional interacting active Brownian particles (ABPs) in which the time variation of the position $\mathbf{r}_i$ and (2D) orientation $\mathbf{q}_i$ of the $i$th particle follow overdamped Langevin equations:
\begin{subequations}
  \label{eq:eom}
    \begin{align}
        \Dot{\mathbf{r}}_i &= U_o \mathbf{q}_i + \frac{1}{\zeta}\sum_{j\neq i}^N\mathbf{F}_{ij} + \bm{\eta}_i\label{eq:rdot}, \\ \Dot{\mathbf{q}}_i &=  \bm{\Omega}_i\times\mathbf{q}_i\label{eq:qdot},
    \end{align} 
\end{subequations}
where $\mathbf{F}_{ij}$ is a pairwise interaction force, $U_o$ is the intrinsic active speed, $\zeta$ is the translational drag coefficient, $\bm{\eta}_i$ is a stochastic translational velocity with zero mean and variance $\langle \bm{\eta}_i (t) \bm{\eta}_j(t')\rangle = 2D_T\delta_{ij}\delta(t-t')\mathbf{I}$, and $\bm{\Omega}_i$ is a stochastic angular velocity.
In two dimensions, the only component of $\bm{\Omega}_i$ relevant to the dynamics of $\mathbf{q}_i$ will be that of the out-of-plane direction, which we describe with unit vector $\mathbf{e}_z$.
We can express $\bm{\Omega}_i$ as $\Omega_i\mathbf{e}_z$, where $\Omega_i$ has zero mean and variance $\langle \Omega_i (t) \Omega_j(t')\rangle = 2D_R\delta_{ij}\delta(t-t')$.
Here, $D_R$ and $D_T$ are the rotational and translational  diffusivity, respectively.
The orientational relaxation time follows as $\tau_R \equiv D_R^{-1}$.
Thus, the characteristic length scale that a non-interacting particle travels before reorienting is given by the intrinsic run length: $\ell_o\equiv U_o\tau_R$.
The pairwise interaction forces are generated by a Weeks-Chandler-Anderson~\cite{Weeks1971} potential with characteristic energy and length scales of $\epsilon$ and $\sigma$, respectively. 

The relative strength of translational noise can be characterized by a dimensionless ``thermalization'' coefficient ${\mathcal{T} \equiv D_T/\left(D_R\sigma^2\right)}$.
Existing simulation studies of ABPs tend to either consider the ``athermal'' limit $\mathcal{T}=0$~\cite{Omar2020,Omar2021,Mallory2021,Langford2024TheoryPhases} or $\mathcal{T} = 1/3$~\cite{Redner2013StructureFluid,Redner2013ReentrantAttraction,Stenhammar2014PhaseDimensionality,Bialke2015,Caporusso2020Motility-InducedSystem,Martin-Roca2021CharacterizationFeatures,Yan2025StochasticMatter}, the latter of which is consistent with the three-dimensional Einstein relation~\cite{Zottl2023ModelingFields} between thermal translation and rotation of an isolated sphere in a viscous medium.
Throughout this Letter, we will refer to simulations with $\mathcal{T} = 0$ as athermal and $\mathcal{T}>0$ as thermal while stressing that $D_T$ (and $D_R$) need not be thermal in origin.
We further emphasize that our particle dynamics remain stochastic and diffusive (at long times) in the athermal limit.

In order to ensure minimal overlap of disks while $\mathcal{T}$ is small, we set the characteristic interaction force to be much larger than the active force with the dimensionless stiffness set to $\mathcal{S}\equiv \epsilon/(\zeta U_o \sigma) = 50$~\cite{Omar2021}.
As detailed in Appendix~\ref{appendix:potential}, our choice of a steep potential allows us to define an effective hard-sphere diameter $d = 2^{1/6}\sigma$ for systems in which the particles are monodisperse in size.
We implement polydisperse simulations by radially shifting the WCA potential and can define the effective hard-disk diameter of the $i$th particle, $d_i$.
We consider a Gaussian distribution of particle diameter with a mean $\langle d_i\rangle = d$ and a standard deviation of $0.09d$ (i.e.,~$9\%)$.
The area fraction occupied by our disks can then be defined as $\phi_o = \rho\pi \langle d_i^2\rangle/4$ where $\rho$ is the areal number density.

Our system state is thus fully described by four dimensionless parameters: $\mathcal{T}$, $\ell_o/\sigma$, $\mathcal{S}$, and $\phi_o$ in addition to the degree of polydispersity~\footnote{We note that the particle activity can also be quantified by defining a P\'{e}clet number $\text{Pe}\equiv U_o\sigma/D_T$.
We can express this P\'{e}clet via our dimensionless run length and thermalization coefficient: $\text{Pe} = \ell_o/\left(\mathcal{T}\sigma\right)$. }.
All simulations are performed at a run length of $\ell_o/\sigma = 112.25$ (or $\ell_o/d = 100$), well beyond the critical activity of 2D MIPS~\cite{Digregorio2018FullSeparation, Omar2023b}.
Monodisperse and polydisperse simulations were conducted at area fractions of $\phi_o=0.64$ and $\phi_o=0.49$, respectively ~\footnote{The significance of this difference in overall area fractions, as well as simulation results for a thermal polydisperse simulation done at $\phi_o=0.64$, is further discussed in the SM.}.
All simulations are performed using the HOOMD-Blue software package~\cite{Anderson2020} and contain a total of $315844$ particles.

\begin{figure}
	\centering
	\includegraphics[width=0.48\textwidth]{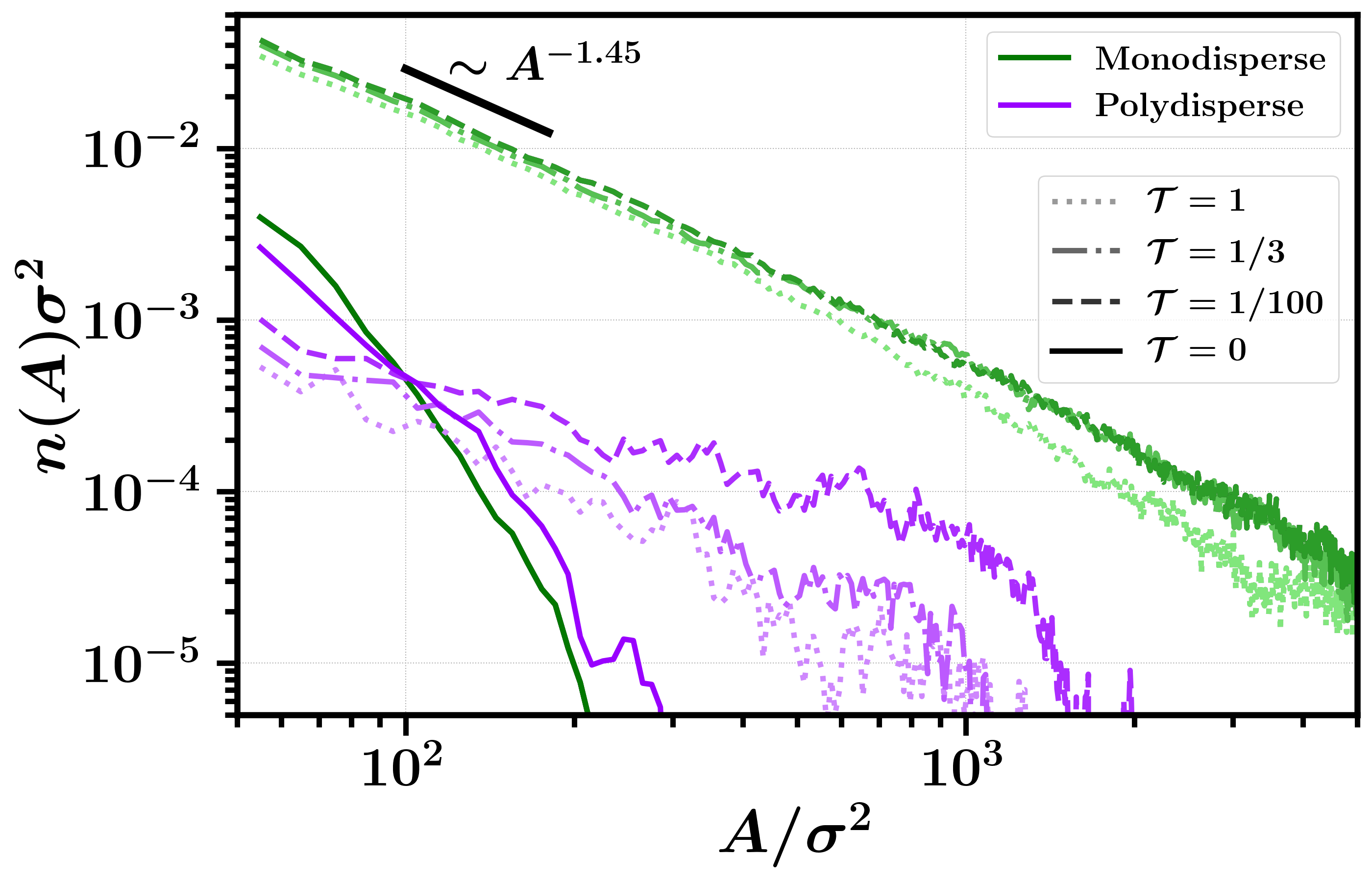}
	\caption{\protect\small{{The dependence of $n(A)$ on thermal noise for both polydisperse and monodisperse systems.}}}
	\label{fig:figure2}
\end{figure}

\textit{Simulation Results.--}
Representative snapshots of the steady-state configurations for both the monodisperse and polydisperse systems are shown in Fig.~\ref{fig:figure1} as a function of $\mathcal{T}$.
Consistent with previous works~\cite{Stenhammar2014PhaseDimensionality,Caporusso2020Motility-InducedSystem,Shi2020Self-OrganizedParticles}, the liquid phase in \textit{monodisperse} and \textit{thermal} systems is decorated with a broad distribution of bubble sizes.
Intriguingly, and particularly for the smallest non-zero value of the thermal noise ($\mathcal{T}=1/100$), the larger bubbles appear to be approaching the size of the vapor phase. 
This may suggest that the liquid in these systems might be in a state of self-organized criticality~\cite{Shi2020Self-OrganizedParticles}.  

One may suspect that $\mathcal{T} = 1/100$ is sufficiently small that we have effectively reached the athermal limit in which translational noise plays a negligible role in the large-length scale phenomenology. 
However, setting the translational noise to be identically zero results in a dramatic change in the bubble distribution: bubbles become extraordinarily rare. 
We can quantify this dramatic change through the bubble distribution, $n(A)$, the average (over time in the steady state) number of bubbles of area $A$~\footnote{$n(A)$ is thus generally an extensive quantity that can be made intensive by, for example, dividing by the area of the liquid phase~\cite{Shi2020Self-OrganizedParticles}.}.
The total number of bubbles (irrespective of size) in the system is thus given by $N_{\rm bub} = \int_0^{\infty} n(A) dA$.
We construct $n(A)$ by calculating the instantaneous coarse-grained density field to identify regions of space with gas phase density, then using the DBSCAN-enabled~\cite{Ester1996ANoise} procedure described in Ref.~\cite{Caporusso2020Motility-InducedSystem} to identify bubbles while ignoring the largest cluster [i.e., the bulk gas phase]. 

The sharp qualitative contrast between the thermal and athermal states are reflected in $n(A)$ [ Fig.~\ref{fig:figure2}]. 
For the thermal monodisperse systems, we find an algebraic decay of $n(A) \sim A^{-1.45}$ that persists for over a decade of area [See Appendix~\ref{appendix:bub}].
This decay cannot continue indefinitely, and eventually we observe a dramatic decrease in bubble number beyond a large cutoff area (see Supplemental Material~\footnote{See Supplemental Material at [URL]}).
For the smallest bubble areas, the athermal system has an order of magnitude fewer bubbles than the thermal systems.
The bubble number for the athermal system sharply decays with area, further widening the difference with the thermal systems.
Bubbles of sizes in excess of one hundred particle areas are essentially improbable for the athermal system while routinely observed for systems with finite translational noise. 

The simulations by Caporusso et al.~\cite{Caporusso2020Motility-InducedSystem} found that bubbles are often observed at the grain boundaries of hexatic domains. 
To control for these effects, we now focus on polydisperse disks and show typical configurations of their steady state in Fig.~\ref{fig:figure1}.
No bubble formation comparable to that of the thermal monodisperse system is observed for either the thermal or athermal polydisperse systems.
$n(A)$ is nearly identical in the athermal polydisperse and monodisperse cases, with larger bubbles completely absent.
As $\mathcal{T}$ is introduced, the polydisperse $n(A)$ broadens and larger bubbles are more likely to be observed, but remain uncommon.

Figure~\ref{fig:figure3}(a) displays the spatial correlation of the per-particle hexatic-order parameter (see Appendix~\ref{appendix:hex}), confirming the absence of hexatic domains in our polydisperse systems, as anticipated. 
For the monodisperse system, hexatic correlations quantitatively increase in spatial range with decreasing translational noise but there is apparently no qualitative distinction. 
This difference in range can be visually appreciated in Figure~\ref{fig:figure3} which colors the distinct domains.
The size distribution can be readily computed for the thermal systems, and appears to be nearly independent of $\mathcal{T}$ [see Fig.~\ref{fig:figure3}(b)]. 
From this analysis, it appears unlikely that the structural differences between the athermal and thermal monodisperse systems are alone responsible for the strong disparity in their bubble statistics.

\begin{figure}
	\centering	\includegraphics[width=0.48\textwidth]{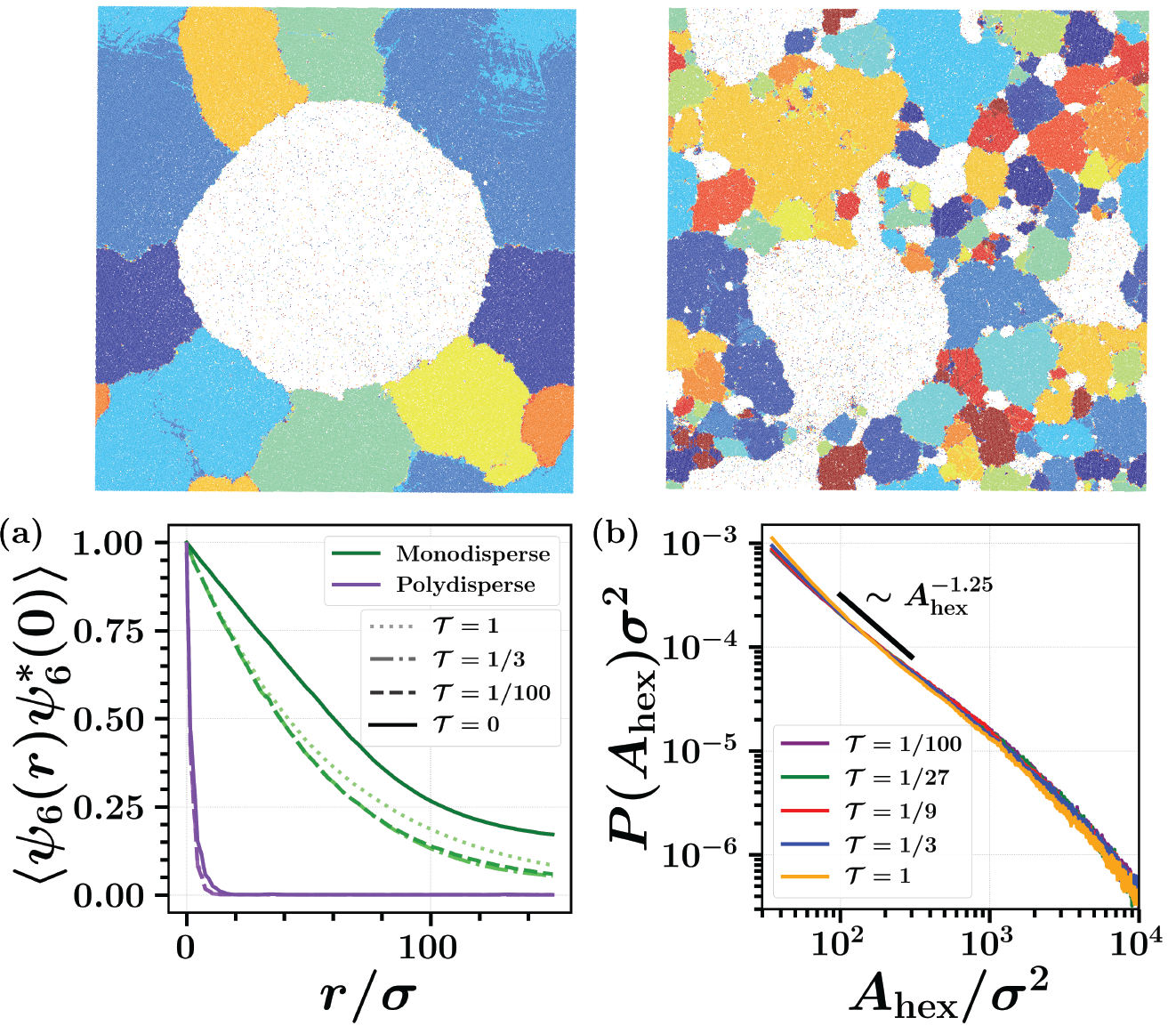}
	\caption{\protect\small{{Top: Snapshots of $\mathcal{T}=0$ (left) and $\mathcal{T}=1/100$ (right) monodisperse simulations colored by hexatic domain orientation, rendered using Ovito~\cite{Stukowski2010VisualizationTool}. (a) Hexatic order parameter spatial correlation functions. (b) Probability distribution of hexatic domain areas, $A_{\rm hex}$, for thermalized monodisperse simulations. Polydisperse simulations not shown due to the lack of domains. Monodisperse $\mathcal{T} = 0$ not shown due to the sluggish kinetics and the presence of only a few large domains leading to poor statistics.}}}
	\label{fig:figure3}
\end{figure}

\begin{figure}
	\centering
	\includegraphics[width=0.48\textwidth]{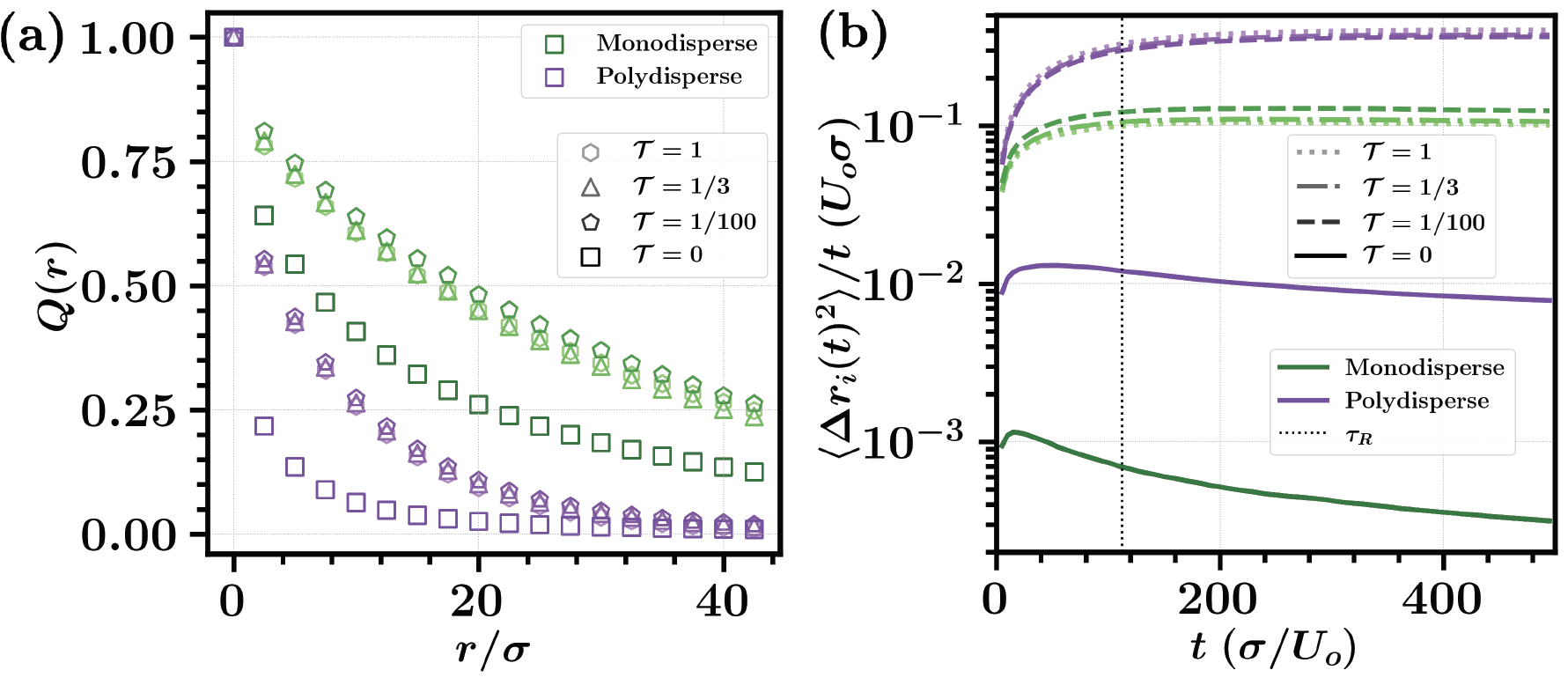}
	\caption{\protect\small{{(a) Spatial correlations of velocity orientation obtained by averaging Eq.~\eqref{eq:caprinicorr} over all particles, employing a time separation $\Delta t$ that maximizes nearest-neighbor correlations. (b) Mean squared displacement of dense phase particles at each simulation condition, normalized by time. Diffusive dynamics should converge to a constant. Vertical dashed line corresponds to $\tau_R$, the timescale beyond which we might expect diffusive dynamics. }}}
	\label{fig:figure4}
\end{figure}

We now look to identify if there are qualitative \textit{dynamical} distinctions which may delineate the thermal monodisperse systems from the athermal and polydisperse systems. 
Direct visualization of the particle dynamics indeed reveals a dependence of the dense-phase particle dynamics on the degree of thermalization and polydispersity (see SM~\cite{Note4} for videos). 
In the athermal simulations, dynamics were sluggish and mediated by point vacancies (in the monodisperse case) or areas of lower density (in the polydisperse case).
Monodisperse thermal simulations displayed striking collective motion, with individual hexatic domains often translating and reorienting as a group.
Polydisperse thermal simulations featured rapid transport of particles without noticeable collective effects.
In order to quantify the collective motion observed in these videos, we calculate the spatial correlation function of the velocity orientation proposed by Caprini \textit{et al.}~\cite{Caprini2020HiddenDisks,Caprini2020SpontaneousSeparation,Caprini2021SpatialParticles}:
\begin{equation}
    Q_i(r) = 1 - 2\sum_{j}^{N_r}\frac{\theta_{ij}}{\mathcal{N}_r\pi},\label{eq:caprinicorr}
\end{equation}
where $\mathcal{N}_{r}$ is the number of particles in a shell of some thickness $\Delta r$ centered at a distance $r$ away from the $i$th particle and $\theta_{ij}$ is the angle between the velocities of particle $i$ and particle $j$. 
Clearly, if the tagged particle velocity is perfectly aligned (anti-aligned) with those of the particles in the shell, we will recover $Q_i(r) = 1$ ($Q_i(r) = -1$). 
We can then calculate the average velocity orientation correlations $Q(r) = \sum_i Q_i(r)/N_{\rm dense}$ by averaging Eq.~\eqref{eq:caprinicorr} over all particles in the dense phase.
The results of this calculation are shown in Fig.~\ref{fig:figure4}(a).
We note that because all simulations are overdamped, the velocities used in this calculation are \textit{average} velocities ${\mathbf{v}_i(t) = (\mathbf{r}_{i}(t+\Delta t)-\mathbf{r}_i(t))/\Delta t}$.
The magnitude of $Q(r)$ therefore depends strongly on the choice of separation time $\Delta t$.
We therefore calculate the correlations of the nearest-neighbor velocity orientations, $Q(r=2^{1/6}\sigma)$, across a wide range of $\Delta t$ (shown in SM~\cite{Note4}) and plot in Fig.~\ref{fig:figure4}(a) the $Q(r)$ corresponding to the $\Delta t$ that maximizes nearest-neighbor correlations at each simulation condition.
While $Q(r)$ allows us to quantify spatial dynamical correlations, we will characterize the rate of transport through the time dependence of the mean squared displacement (MSD), displayed in Fig.~\ref{fig:figure4}(b).

The spatial range of velocity correlations increases with thermal noise in both the monodisperse and polydisperse simulations.
All monodisperse simulations had longer range velocity correlations, likely due to the longer range hexatic ordering.
Any finite amount of thermalization, for both monodisperse and polydisperse simulations, increased the MSD of particles in the dense phase by multiple orders of magnitude.
We note that our averaging procedure includes only dense-phase particles and excludes particles that escaped the dense phase during our observation window. 
This increase in both the range of dynamical correlations as well as the average speed of transport with $\mathcal{T}$ points to thermalization triggering some form of cooperative motion.

\textit{Discussion.--}
The strong dependence of any measured quantity on thermalization is striking.
Consider the relative magnitudes of terms in Eq.~\eqref{eq:rdot}.
If we consider the effective diffusivity of an ideal ABP, it would be described by~\cite{Takatori2014}:
\begin{equation}
    D_{\rm ideal} = \frac{U_o^2}{2 D_R} + D_T = \frac{U_o \ell_o}{2} + \frac{\mathcal{T}\sigma^2}{\ell_o} \label{eq:idealdiffusivity}.
\end{equation}
Therefore, at the simulated run length of $\ell_o = 112.25\sigma$, the thermal contribution to the mobility of an ideal particle is negligible even as high as $\mathcal{T} = 1$.
While it is true that the active contribution to this ideal diffusivity should be far lower in the dense phase of interacting ABPs, the dramatic increase in bubble fraction and particle mobility with thermalization still exists at values of $\mathcal{T}$ as low as $1/100$. 
Capturing these effects in a coarse-grained or continuum-level model may warrant further investigation into cooperative effects that require consideration of long-range correlations between particle dynamics that are not considered in existing efforts towards coarse-graining ABPs~\cite{Vrugt2023HowTutorial,Langford2025TheMatter}.

Recently, Fausti \textit{et al.}~\cite{Fausti2024StatisticalFluids} argued that $n(A)$ should be understood as the net result of several processes including nucleation, ripening/coarsening, and coalescence. 
They also proposed a reduced model which allowed one to independently vary the relative strengths of coalescence and (reversed Ostwald) ripening while coupling the nucleation rate to the rate of coalescence. 
From these reduced models one can measure $n(A)$ to be strongly peaked at a particular area when reversed Ostwald ripening dominates or power-law distributed when coalescence/nucleation dominate.
This reduced model may be a crucial tool for qualitative mechanistic insight on the $n(A)$ measured both in this study as well as previous studies~\cite{Caporusso2020Motility-InducedSystem,Shi2020Self-OrganizedParticles,Yan2025StochasticMatter}.
Future work could focus on expanding these models to encompass a wider range of qualitative behaviors so that they are more readily applicable to interpreting simulation data.
For example, the model reported in Ref.~\cite{Fausti2024StatisticalFluids} has no mechanism for a bubble smaller than all of its neighbors to shrink in size, which is difficult to reconcile with the findings of Ref.~\cite{Yan2025StochasticMatter} that the overwhelming majority of small bubbles dissolve into the dense phase.
Other lines of expansion to the model may be to allow for forward Ostwald ripening or an uncoupling of the nucleation rate from the coalescence rate (while still maintaining mass conservation). 

\begin{figure}
	\centering
	\includegraphics[width=0.48\textwidth]{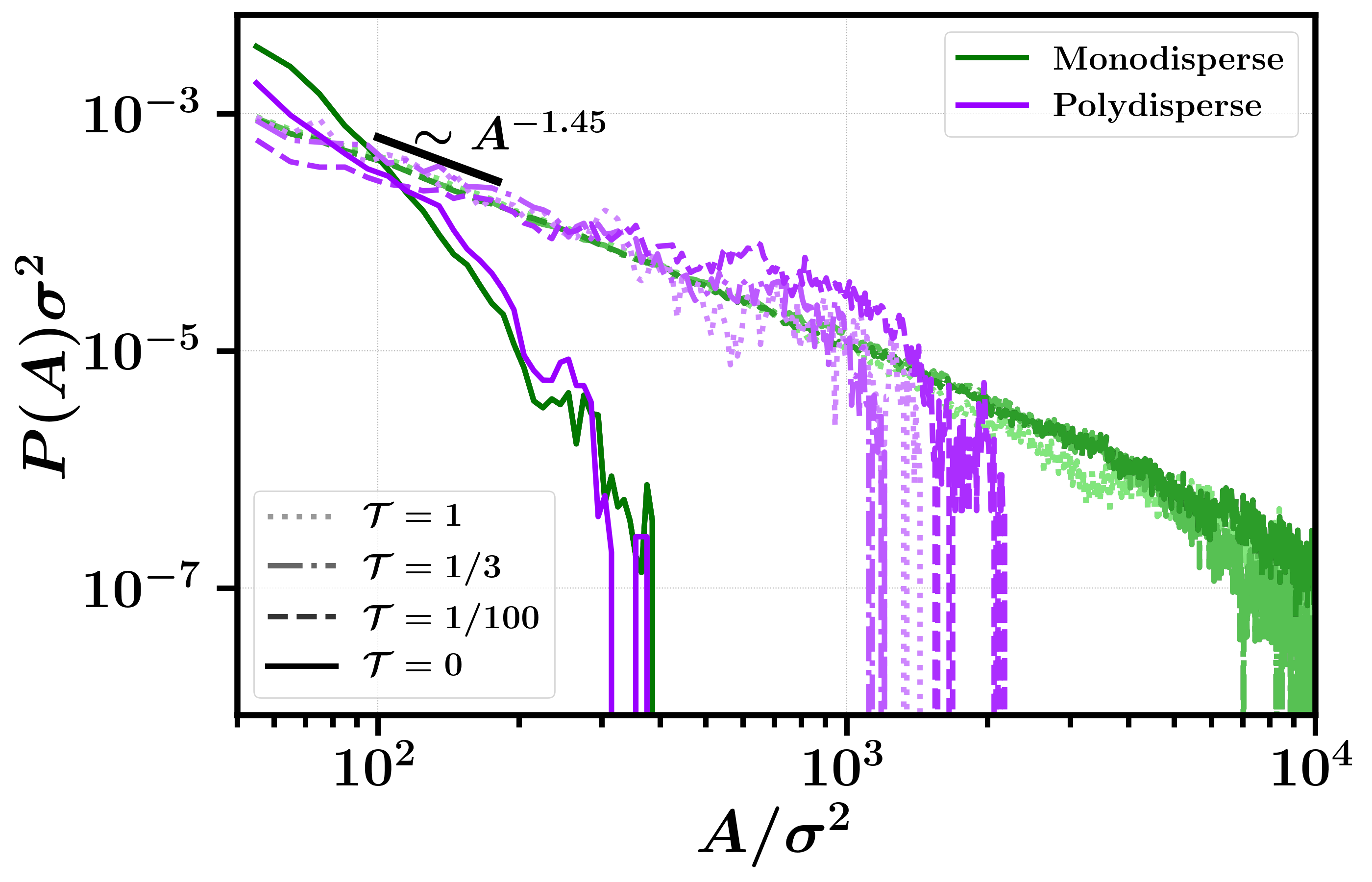}
	\caption{\protect\small{{Probability distribution of observing a bubble of size $A$ at various simulation conditions. Although the thermalized polydisperse systems are noisy due to the scarcity of bubbles, the scaling appears similar to the thermalized monodisperse systems.}}}
	\label{fig:figure5}
\end{figure}

We can further isolate the importance of translational noise by defining $P(A)$, the probability distribution of any bubble being of size $A$.
The size distribution can be found by normalizing $n(A)$ by its zeroth moment, $N_{\rm bub}$.
$P(A)$ [shown in Fig.~\ref{fig:figure5}] thus solely measures the size distribution, containing no information on the number of bubbles.
Although the data remains noisy for systems in which bubbles were rare, all thermalized systems (including polydisperse) seem to display the same algebraic decay for small bubble sizes.
This might suggest the possibility that the structural and dynamical distinctions between the polydisperse and monodisperse systems solely impact $N_{\rm bub}$ while giving rise to the same size distribution.
Importantly, $P(A)$ for the \textit{athermal} monodisperse and polydisperse systems does not collapse onto those of the thermal systems. 

Hexatic order has a significant impact on the \textit{number of bubbles} formed in the steady state but appears to be less decisive for the size distribution.   
This may explain the lack of bubbles in two-dimensional ABPs, since the liquid phase in three dimensions is disordered~\cite{Omar2021}.
The apparent crucial role of hexatic order was first appreciated by Caporusso \textit{et al.}~\cite{Caporusso2020Motility-InducedSystem}, who identified that bubbles often appear at grain boundaries.
It may be that these boundaries serve as preferred nucleation sites~\cite{Cates2025ActivePhysics} in comparison to dense liquids (polydisperse) or ordered (monodisperse) domains.
We note, however, that the fraction of gas phase within bubbles (see Appendix~\ref{appendix:bub}) indicates that the number of observed bubbles decreases with $\mathcal{T}$ while $\mathcal{T}>0$ (see also Fig.~\ref{fig:figure1}), despite the hexatic domain structure of monodisperse thermal simulations being indistinguishable (see Fig.~\ref{fig:figure3}).

\textit{Conclusions.-- }
In summary, we conducted simulations of phase-separated monodisperse and polydisperse ABPs at varying degrees of thermalization.
We found that only monodisperse ABPs with finite (no matter how small) thermalization were well-endowed with bubbles in the dense phase.
We also found that thermalization resulted in dense phase dynamics that were both faster and longer-range in spatial correlations, suggesting that translational noise may induce cooperative motion. 
These findings present data that further complicate an ongoing debate regarding the origin of bubbles in two-dimensional active matter.
Further work, both theoretical and numerical, is needed to understand the drastic dependence of both the bubble population and the dense phase dynamics in two-dimensional ABPs.

\begin{acknowledgments}
\textit{Acknowledgments.-- }
This material is based upon work supported by the U.S. Department of Energy, Office of Science, Basic Energy Sciences (BES) under Award Number DE-SC0024900. 
L.L. was supported in part by the Department of Defense (DoD) through the National Defense Science \& Engineering Graduate (NDSEG) Fellowship Program. 
The authors also thank David King, Eric Weiner, Yilong Zhou, and Daniel Evans for insightful discussions and suggestions.
\end{acknowledgments}

\section{End Matter}

\fakesection{Appendix I: Potential Details}\label{appendix:potential}
\textit{Appendix I: Potential Details.--}
The pairwise force between particles in Eq.~\eqref{eq:eom} can be written as the gradient of a two-body interaction potential, i.e.,~$\mathbf{F}_{ij} = -\boldsymbol{\nabla} V_{ij}(|\mathbf{r}_i-\mathbf{r}_j|)$, where $\boldsymbol{\nabla}$ is the gradient with respect to the vector connecting particle $i$ and $j$.
In all simulations we employ a radially shifted Lennard-Jones potential~\cite{Phillips2010StabilitySystems} of the form:
\begin{equation}
V(r_{ij})/\epsilon =
 \begin{cases}
   4\left[\left(\frac{\sigma}{r_{ij}-\alpha_{ij}}\right)^{12} - \left(\frac{\sigma}{r_{ij}-\alpha_{ij}}\right)^{6}\right] + 1, & r_{ij} \le r^{\rm cut}_{ij} \\
    0, & r_{ij} > r^{\rm cut}_{ij}.  \\
    \end{cases}
\end{equation}
where $r_{ij} = |\mathbf{r}_i-\mathbf{r}_j|$, $\epsilon$ is the characteristic Lennard-Jones (LJ) energy,  $\sigma$ is the LJ diameter, and $\alpha_{ij}$ is the radial shift. 
The potential is cut off at the minimum, ensuring purely repulsive interactions with ${r^{\rm cut}_{ij} \equiv 2^{1/6}\sigma + \alpha_{ij}}$.
In the absence of a radial shift, our potential reduces to the WCA potential~\cite{Weeks1971} and the cutoff distance can be interpreted as an effective hard-disk diameter (for sufficiently stiff potentials) with ${r_{\rm cut} = 2^{1/6}\sigma \equiv d}$.
We allow particles to differ in size through $\alpha_{ij}$: a pair of particles with distinct diameters begin to feel repulsion at ${r^{\rm cut}_{ij}=\alpha_{ij} + d}$ which we take to be the mean particle diameter of the pair (i.e.,~${d_{ij} = (d_i+d_j)/2}$ where $d_i$ is the hard-disk diameter of particle $i$).

For the athermal ($\mathcal{T} = 0$) simulations, particles were initialized randomly and the simulations were continued until the system reached a fully coarsened steady state with a single liquid domain and a single gas domain.
In the polydisperse simulation, this was done using a square $800\sigma \times 800\sigma$ simulation box, resulting in an overall area fraction of $\phi_o$ of approximately $0.49$.
The kinetics of the monodisperse case were sluggish, and achieving a final state with a single liquid and gas domain within reasonable simulation time required using a smaller $700\sigma \times 700\sigma$ simulation box, resulting in an overall area fraction of $\phi_o$ of approximately $0.64$. 
Despite the difference in overall area fraction, both athermal simulations reached qualitatively similar steady states.
After reaching a steady state, statistics were then collected over a minimum time span of $5\times 10^4 \sigma/U_o$. 
All thermal simulations ($\mathcal{T} > 0$) were initialized from the final simulation frame of the athermal simulations, allowed to reach a steady state over $5\times 10^4 \sigma/U_o$, then integrated for an additional $5\times 10^4 \sigma/U_o$ in order to collect statistics.
The monodisperse simulations were conducted at thermalization coefficients of $\mathcal{T} = \{0,\frac{1}{100},\frac{1}{27},\frac{1}{9},\frac{1}{3},1\}$ while polydisperse simulations were calculated at thermalization coefficients of $\mathcal{T} = \{0,\frac{1}{100},\frac{1}{3},1\}$.
In order to avoid appreciably shifting the critical point~\cite{Note4} (and therefore the phase diagram), we restrict our simulations to thermalization coefficients $\mathcal{T} \leq 1$.
\fakesection{Appendix II: Statistical Properties of Bubbles}\label{appendix:bub}
\textit{Appendix II: Statistical Properties of Bubbles.--}
$n(A)$ was calculated by evaluating the coarse-grained density field to identify patches of space where the local density corresponded to the gas phase.
The DBSCAN algorithm was then used to identify clusters of gas phase which were identified as bubbles, with the largest cluster (corresponding to the bulk gas phase) ignored.
Further details are provided in the Supplemental Material~\cite{Note4}.

$n(A)$ measured from thermalized monodisperse simulations is well described by a power-law, in agreement with Refs.~\cite{Caporusso2020Motility-InducedSystem,Shi2020Self-OrganizedParticles}.
However, in all other simulations where bubbles are rarefied, the statistical convergence is poor and it is less clear what functional form is appropriate.
We therefore fit $n(A)$ from only the monodisperse thermal simulations to a power law $n(A)\sim A^{-\omega}$ and report $\omega$ in Table~\ref{tab:bubbles}. 
By identifying the total area occupied by bubbles $A_{\rm bub}= \int An(A)dA $ and the bulk gas phase $A_{\rm gas}$, we can define the fraction of total gas area occupied by bubbles as $\phi_B = A_{\rm bub}/(A_{\rm bub} + A_{\rm gas})$.
We report $\phi_B$ in Table~\ref{tab:bubbles}, where we can see that the thermalized monodisperse simulations have $\phi_B$ nearly three orders of magnitude larger than the other simulation conditions.

\begin{table}
\caption{\label{tab:bubbles} Scaling exponents of $n(A)$ and area fraction of gas phase in bubbles, $\phi_B$}
\begin{ruledtabular}
\begin{tabular}{lccc}
\toprule
Dispersity & $\mathcal{T}$ & $\omega$ & $\phi_B$ \\
\midrule
Monodisperse & $1$ & $1.44\pm 0.08$ & $0.12\pm0.03$\\
Monodisperse & $1/3$ & $ 1.45\pm0.07$ & $0.15\pm0.03$ \\
Monodisperse & $1/9$ & $1.47\pm0.07$ & $0.21\pm0.05$ \\
Monodisperse & $1/27$ & $1.46\pm0.05$ & $0.21\pm0.05$ \\
Monodisperse & $1/100$ & $1.44\pm0.05$ & $0.28\pm0.07$ \\
Monodisperse & $0$ & N/A & $(5.2\pm2.7)\times10^{-4}$ \\
Polydisperse & $1$ & N/A & $(7.1\pm9.9)\times10^{-4}$\\
Polydisperse & $1/3$ & N/A & $(8.0\pm6.6)\times10^{-4}$ \\
Polydisperse & $1/100$ & N/A & $(2.7\pm1.6)\times10^{-3}$ \\
Polydisperse & $0$ & N/A & $(3.3\pm2.2)\times10^{-4}$ \\
\bottomrule
\end{tabular}
\end{ruledtabular}
\end{table}

\fakesection{Appendix III: Statistical Properties of Hexatic Domains}\label{appendix:hex}
\textit{Appendix III: Statistical Properties of Hexatic Domains.--}
The hexatic order parameter of each particle is ${\psi_{6j} = N_j^{-1}\sum_{k=1}^{N_j}\text{exp}[i6\phi_{jk}]}$, where $N_j$ is the number of neighbors of particle $j$ and $\phi_{jk}$ is the angle between the vector connecting particles $j$ and $k$ with an arbitrary reference vector.
This calculation was facilitated using the Freud analysis library~\cite{Ramasubramani2020Freud:Data}.
From the per-particle hexatic order parameter, we compute the correlation function $\langle \psi_{6}(r)\psi_{6j}^*\rangle$ where $\psi_6(r)$ is the average hexatic order parameter of particles a distance $r$ away from particle $j$ and the angled brackets indicate an average over all particles within the dense phase.
The results of this calculation are plotted in Fig.~\ref{fig:figure3}(a).
We note that the system does \textit{not} have translational symmetry as it is phase separated, so this correlation function will be influenced by particles at the interface of the dense and gas phases. 
Regardless, all polydisperse simulations had hexatic correlations decay within a distance on the order of $\sigma$ while the monodisperse simulations had correlations indicative of larger hexatic domains [see Fig.~\ref{fig:figure3}(a) and (b)]. 
The distribution of sizes for these hexatic domains was calculated following Ref.~\cite{Caporusso2020Motility-InducedSystem} by calculating the phase angle of $\psi_{6j}$ for each particle.
Clusters of particles with similar phase angles were identified as hexatic domains, which corresponds to the coloration in Fig.~\ref{fig:figure3}, and the probability distribution of finding hexatic domains of a particular area is shown in Fig.~\ref{fig:figure3}(b).


\begin{thebibliography}{52}%
\makeatletter
\providecommand \@ifxundefined [1]{%
 \@ifx{#1\undefined}
}%
\providecommand \@ifnum [1]{%
 \ifnum #1\expandafter \@firstoftwo
 \else \expandafter \@secondoftwo
 \fi
}%
\providecommand \@ifx [1]{%
 \ifx #1\expandafter \@firstoftwo
 \else \expandafter \@secondoftwo
 \fi
}%
\providecommand \natexlab [1]{#1}%
\providecommand \enquote  [1]{``#1''}%
\providecommand \bibnamefont  [1]{#1}%
\providecommand \bibfnamefont [1]{#1}%
\providecommand \citenamefont [1]{#1}%
\providecommand \href@noop [0]{\@secondoftwo}%
\providecommand \href [0]{\begingroup \@sanitize@url \@href}%
\providecommand \@href[1]{\@@startlink{#1}\@@href}%
\providecommand \@@href[1]{\endgroup#1\@@endlink}%
\providecommand \@sanitize@url [0]{\catcode `\\12\catcode `\$12\catcode `\&12\catcode `\#12\catcode `\^12\catcode `\_12\catcode `\%12\relax}%
\providecommand \@@startlink[1]{}%
\providecommand \@@endlink[0]{}%
\providecommand \url  [0]{\begingroup\@sanitize@url \@url }%
\providecommand \@url [1]{\endgroup\@href {#1}{\urlprefix }}%
\providecommand \urlprefix  [0]{URL }%
\providecommand \Eprint [0]{\href }%
\providecommand \doibase [0]{https://doi.org/}%
\providecommand \selectlanguage [0]{\@gobble}%
\providecommand \bibinfo  [0]{\@secondoftwo}%
\providecommand \bibfield  [0]{\@secondoftwo}%
\providecommand \translation [1]{[#1]}%
\providecommand \BibitemOpen [0]{}%
\providecommand \bibitemStop [0]{}%
\providecommand \bibitemNoStop [0]{.\EOS\space}%
\providecommand \EOS [0]{\spacefactor3000\relax}%
\providecommand \BibitemShut  [1]{\csname bibitem#1\endcsname}%
\let\auto@bib@innerbib\@empty
\bibitem [{\citenamefont {Brangwynne}\ \emph {et~al.}(2009)\citenamefont {Brangwynne}, \citenamefont {Eckmann}, \citenamefont {Courson}, \citenamefont {Rybarska}, \citenamefont {Hoege}, \citenamefont {Gharakhani}, \citenamefont {J{\"{u}}licher},\ and\ \citenamefont {Hyman}}]{Brangwynne2009GermlineDissolution/Condensation}%
  \BibitemOpen
  \bibfield  {author} {\bibinfo {author} {\bibfnamefont {C.~P.}\ \bibnamefont {Brangwynne}}, \bibinfo {author} {\bibfnamefont {C.~R.}\ \bibnamefont {Eckmann}}, \bibinfo {author} {\bibfnamefont {D.~S.}\ \bibnamefont {Courson}}, \bibinfo {author} {\bibfnamefont {A.}~\bibnamefont {Rybarska}}, \bibinfo {author} {\bibfnamefont {C.}~\bibnamefont {Hoege}}, \bibinfo {author} {\bibfnamefont {J.}~\bibnamefont {Gharakhani}}, \bibinfo {author} {\bibfnamefont {F.}~\bibnamefont {J{\"{u}}licher}},\ and\ \bibinfo {author} {\bibfnamefont {A.~A.}\ \bibnamefont {Hyman}},\ }\href {https://doi.org/10.1126/science.1172046} {\bibfield  {journal} {\bibinfo  {journal} {Science}\ }\textbf {\bibinfo {volume} {324}},\ \bibinfo {pages} {1729} (\bibinfo {year} {2009})}\BibitemShut {NoStop}%
\bibitem [{\citenamefont {Palacci}\ \emph {et~al.}(2013)\citenamefont {Palacci}, \citenamefont {Sacanna}, \citenamefont {Steinberg}, \citenamefont {Pine},\ and\ \citenamefont {Chaikin}}]{Palacci2013LivingSurfers}%
  \BibitemOpen
  \bibfield  {author} {\bibinfo {author} {\bibfnamefont {J.}~\bibnamefont {Palacci}}, \bibinfo {author} {\bibfnamefont {S.}~\bibnamefont {Sacanna}}, \bibinfo {author} {\bibfnamefont {A.~P.}\ \bibnamefont {Steinberg}}, \bibinfo {author} {\bibfnamefont {D.~J.}\ \bibnamefont {Pine}},\ and\ \bibinfo {author} {\bibfnamefont {P.~M.}\ \bibnamefont {Chaikin}},\ }\href {https://doi.org/10.1126/science.1230020} {\bibfield  {journal} {\bibinfo  {journal} {Science}\ }\textbf {\bibinfo {volume} {339}},\ \bibinfo {pages} {936} (\bibinfo {year} {2013})}\BibitemShut {NoStop}%
\bibitem [{\citenamefont {Han}\ \emph {et~al.}(2017)\citenamefont {Han}, \citenamefont {Yan}, \citenamefont {Granick},\ and\ \citenamefont {Luijten}}]{Han2017EffectiveMixture}%
  \BibitemOpen
  \bibfield  {author} {\bibinfo {author} {\bibfnamefont {M.}~\bibnamefont {Han}}, \bibinfo {author} {\bibfnamefont {J.}~\bibnamefont {Yan}}, \bibinfo {author} {\bibfnamefont {S.}~\bibnamefont {Granick}},\ and\ \bibinfo {author} {\bibfnamefont {E.}~\bibnamefont {Luijten}},\ }\href {https://doi.org/10.1073/pnas.1706702114} {\bibfield  {journal} {\bibinfo  {journal} {Proceedings of the National Academy of Sciences}\ }\textbf {\bibinfo {volume} {114}},\ \bibinfo {pages} {7513} (\bibinfo {year} {2017})}\BibitemShut {NoStop}%
\bibitem [{\citenamefont {Liu}\ \emph {et~al.}(2019)\citenamefont {Liu}, \citenamefont {Patch}, \citenamefont {Bahar}, \citenamefont {Yllanes}, \citenamefont {Welch}, \citenamefont {Marchetti}, \citenamefont {Thutupalli},\ and\ \citenamefont {Shaevitz}}]{Liu2019Self-DrivenFormation}%
  \BibitemOpen
  \bibfield  {author} {\bibinfo {author} {\bibfnamefont {G.}~\bibnamefont {Liu}}, \bibinfo {author} {\bibfnamefont {A.}~\bibnamefont {Patch}}, \bibinfo {author} {\bibfnamefont {F.}~\bibnamefont {Bahar}}, \bibinfo {author} {\bibfnamefont {D.}~\bibnamefont {Yllanes}}, \bibinfo {author} {\bibfnamefont {R.~D.}\ \bibnamefont {Welch}}, \bibinfo {author} {\bibfnamefont {M.~C.}\ \bibnamefont {Marchetti}}, \bibinfo {author} {\bibfnamefont {S.}~\bibnamefont {Thutupalli}},\ and\ \bibinfo {author} {\bibfnamefont {J.~W.}\ \bibnamefont {Shaevitz}},\ }\href {https://doi.org/10.1103/PhysRevLett.122.248102} {\bibfield  {journal} {\bibinfo  {journal} {Phys. Rev. Lett.}\ }\textbf {\bibinfo {volume} {122}},\ \bibinfo {pages} {248102} (\bibinfo {year} {2019})}\BibitemShut {NoStop}%
\bibitem [{\citenamefont {Soni}\ \emph {et~al.}(2019)\citenamefont {Soni}, \citenamefont {Bililign}, \citenamefont {Magkiriadou}, \citenamefont {Sacanna}, \citenamefont {Bartolo}, \citenamefont {Shelley},\ and\ \citenamefont {Irvine}}]{Soni2019TheFluid}%
  \BibitemOpen
  \bibfield  {author} {\bibinfo {author} {\bibfnamefont {V.}~\bibnamefont {Soni}}, \bibinfo {author} {\bibfnamefont {E.~S.}\ \bibnamefont {Bililign}}, \bibinfo {author} {\bibfnamefont {S.}~\bibnamefont {Magkiriadou}}, \bibinfo {author} {\bibfnamefont {S.}~\bibnamefont {Sacanna}}, \bibinfo {author} {\bibfnamefont {D.}~\bibnamefont {Bartolo}}, \bibinfo {author} {\bibfnamefont {M.~J.}\ \bibnamefont {Shelley}},\ and\ \bibinfo {author} {\bibfnamefont {W.~T.~M.}\ \bibnamefont {Irvine}},\ }\href {https://doi.org/10.1038/s41567-019-0603-8} {\bibfield  {journal} {\bibinfo  {journal} {Nat. Phys.}\ }\textbf {\bibinfo {volume} {15}},\ \bibinfo {pages} {1188} (\bibinfo {year} {2019})}\BibitemShut {NoStop}%
\bibitem [{\citenamefont {Geyer}\ \emph {et~al.}(2019)\citenamefont {Geyer}, \citenamefont {Martin}, \citenamefont {Tailleur},\ and\ \citenamefont {Bartolo}}]{Geyer2019FreezingLiquids}%
  \BibitemOpen
  \bibfield  {author} {\bibinfo {author} {\bibfnamefont {D.}~\bibnamefont {Geyer}}, \bibinfo {author} {\bibfnamefont {D.}~\bibnamefont {Martin}}, \bibinfo {author} {\bibfnamefont {J.}~\bibnamefont {Tailleur}},\ and\ \bibinfo {author} {\bibfnamefont {D.}~\bibnamefont {Bartolo}},\ }\href {https://doi.org/10.1103/PhysRevX.9.031043} {\bibfield  {journal} {\bibinfo  {journal} {Physical Review X}\ }\textbf {\bibinfo {volume} {9}},\ \bibinfo {pages} {031043} (\bibinfo {year} {2019})}\BibitemShut {NoStop}%
\bibitem [{\citenamefont {Zhang}\ \emph {et~al.}(2021)\citenamefont {Zhang}, \citenamefont {Alert}, \citenamefont {Yan}, \citenamefont {Wingreen},\ and\ \citenamefont {Granick}}]{Zhang2021ActiveDensity}%
  \BibitemOpen
  \bibfield  {author} {\bibinfo {author} {\bibfnamefont {J.}~\bibnamefont {Zhang}}, \bibinfo {author} {\bibfnamefont {R.}~\bibnamefont {Alert}}, \bibinfo {author} {\bibfnamefont {J.}~\bibnamefont {Yan}}, \bibinfo {author} {\bibfnamefont {N.~S.}\ \bibnamefont {Wingreen}},\ and\ \bibinfo {author} {\bibfnamefont {S.}~\bibnamefont {Granick}},\ }\href {https://doi.org/10.1038/s41567-021-01238-8} {\bibfield  {journal} {\bibinfo  {journal} {Nat. Phys.}\ }\textbf {\bibinfo {volume} {17}},\ \bibinfo {pages} {961} (\bibinfo {year} {2021})}\BibitemShut {NoStop}%
\bibitem [{\citenamefont {Cates}\ and\ \citenamefont {Tailleur}(2015)}]{Cates2015}%
  \BibitemOpen
  \bibfield  {author} {\bibinfo {author} {\bibfnamefont {M.~E.}\ \bibnamefont {Cates}}\ and\ \bibinfo {author} {\bibfnamefont {J.}~\bibnamefont {Tailleur}},\ }\href {https://www.annualreviews.org/doi/abs/10.1146/annurev-conmatphys-031214-014710} {\bibfield  {journal} {\bibinfo  {journal} {Annu. Rev. Condens. Matter Phys.}\ }\textbf {\bibinfo {volume} {6}},\ \bibinfo {pages} {219} (\bibinfo {year} {2015})}\BibitemShut {NoStop}%
\bibitem [{\citenamefont {Wittkowski}\ \emph {et~al.}(2014)\citenamefont {Wittkowski}, \citenamefont {Tiribocchi}, \citenamefont {Stenhammar}, \citenamefont {Allen}, \citenamefont {Marenduzzo},\ and\ \citenamefont {Cates}}]{Wittkowski2014ScalarSeparation}%
  \BibitemOpen
  \bibfield  {author} {\bibinfo {author} {\bibfnamefont {R.}~\bibnamefont {Wittkowski}}, \bibinfo {author} {\bibfnamefont {A.}~\bibnamefont {Tiribocchi}}, \bibinfo {author} {\bibfnamefont {J.}~\bibnamefont {Stenhammar}}, \bibinfo {author} {\bibfnamefont {R.~J.}\ \bibnamefont {Allen}}, \bibinfo {author} {\bibfnamefont {D.}~\bibnamefont {Marenduzzo}},\ and\ \bibinfo {author} {\bibfnamefont {M.~E.}\ \bibnamefont {Cates}},\ }\href {https://doi.org/10.1038/ncomms5351} {\bibfield  {journal} {\bibinfo  {journal} {Nat. Commun.}\ }\textbf {\bibinfo {volume} {5}},\ \bibinfo {pages} {4351} (\bibinfo {year} {2014})}\BibitemShut {NoStop}%
\bibitem [{\citenamefont {Tjhung}\ \emph {et~al.}(2018)\citenamefont {Tjhung}, \citenamefont {Nardini},\ and\ \citenamefont {Cates}}]{Tjhung2018}%
  \BibitemOpen
  \bibfield  {author} {\bibinfo {author} {\bibfnamefont {E.}~\bibnamefont {Tjhung}}, \bibinfo {author} {\bibfnamefont {C.}~\bibnamefont {Nardini}},\ and\ \bibinfo {author} {\bibfnamefont {M.~E.}\ \bibnamefont {Cates}},\ }\href {https://link.aps.org/doi/10.1103/PhysRevX.8.031080} {\bibfield  {journal} {\bibinfo  {journal} {Phys. Rev. X}\ }\textbf {\bibinfo {volume} {8}},\ \bibinfo {pages} {031080} (\bibinfo {year} {2018})}\BibitemShut {NoStop}%
\bibitem [{\citenamefont {Fily}\ and\ \citenamefont {Marchetti}(2012)}]{Fily2012}%
  \BibitemOpen
  \bibfield  {author} {\bibinfo {author} {\bibfnamefont {Y.}~\bibnamefont {Fily}}\ and\ \bibinfo {author} {\bibfnamefont {M.~C.}\ \bibnamefont {Marchetti}},\ }\href {https://journals.aps.org/prl/abstract/10.1103/PhysRevLett.108.235702} {\bibfield  {journal} {\bibinfo  {journal} {Phys. Rev. Lett.}\ }\textbf {\bibinfo {volume} {108}},\ \bibinfo {pages} {235702} (\bibinfo {year} {2012})}\BibitemShut {NoStop}%
\bibitem [{\citenamefont {Redner}\ \emph {et~al.}(2013{\natexlab{a}})\citenamefont {Redner}, \citenamefont {Hagan},\ and\ \citenamefont {Baskaran}}]{Redner2013StructureFluid}%
  \BibitemOpen
  \bibfield  {author} {\bibinfo {author} {\bibfnamefont {G.~S.}\ \bibnamefont {Redner}}, \bibinfo {author} {\bibfnamefont {M.~F.}\ \bibnamefont {Hagan}},\ and\ \bibinfo {author} {\bibfnamefont {A.}~\bibnamefont {Baskaran}},\ }\href {https://doi.org/10.1103/PhysRevLett.110.055701} {\bibfield  {journal} {\bibinfo  {journal} {Phy. Rev. Lett.}\ }\textbf {\bibinfo {volume} {110}},\ \bibinfo {pages} {055701} (\bibinfo {year} {2013}{\natexlab{a}})}\BibitemShut {NoStop}%
\bibitem [{\citenamefont {Stenhammar}\ \emph {et~al.}(2014)\citenamefont {Stenhammar}, \citenamefont {Marenduzzo}, \citenamefont {Allen},\ and\ \citenamefont {Cates}}]{Stenhammar2014PhaseDimensionality}%
  \BibitemOpen
  \bibfield  {author} {\bibinfo {author} {\bibfnamefont {J.}~\bibnamefont {Stenhammar}}, \bibinfo {author} {\bibfnamefont {D.}~\bibnamefont {Marenduzzo}}, \bibinfo {author} {\bibfnamefont {R.~J.}\ \bibnamefont {Allen}},\ and\ \bibinfo {author} {\bibfnamefont {M.~E.}\ \bibnamefont {Cates}},\ }\href {https://doi.org/10.1039/C3SM52813H} {\bibfield  {journal} {\bibinfo  {journal} {Soft Matter}\ }\textbf {\bibinfo {volume} {10}},\ \bibinfo {pages} {1489} (\bibinfo {year} {2014})}\BibitemShut {NoStop}%
\bibitem [{\citenamefont {Solon}\ \emph {et~al.}(2018{\natexlab{a}})\citenamefont {Solon}, \citenamefont {Stenhammar}, \citenamefont {Cates}, \citenamefont {Kafri},\ and\ \citenamefont {Tailleur}}]{Solon2018GeneralizedEnsembles}%
  \BibitemOpen
  \bibfield  {author} {\bibinfo {author} {\bibfnamefont {A.~P.}\ \bibnamefont {Solon}}, \bibinfo {author} {\bibfnamefont {J.}~\bibnamefont {Stenhammar}}, \bibinfo {author} {\bibfnamefont {M.~E.}\ \bibnamefont {Cates}}, \bibinfo {author} {\bibfnamefont {Y.}~\bibnamefont {Kafri}},\ and\ \bibinfo {author} {\bibfnamefont {J.}~\bibnamefont {Tailleur}},\ }\href {https://dx.doi.org/10.1088/1367-2630/aaccdd} {\bibfield  {journal} {\bibinfo  {journal} {New J. Phys.}\ }\textbf {\bibinfo {volume} {20}},\ \bibinfo {pages} {075001} (\bibinfo {year} {2018}{\natexlab{a}})}\BibitemShut {NoStop}%
\bibitem [{\citenamefont {Solon}\ \emph {et~al.}(2018{\natexlab{b}})\citenamefont {Solon}, \citenamefont {Stenhammar}, \citenamefont {Cates}, \citenamefont {Kafri},\ and\ \citenamefont {Tailleur}}]{Solon2018GeneralizedMatter}%
  \BibitemOpen
  \bibfield  {author} {\bibinfo {author} {\bibfnamefont {A.~P.}\ \bibnamefont {Solon}}, \bibinfo {author} {\bibfnamefont {J.}~\bibnamefont {Stenhammar}}, \bibinfo {author} {\bibfnamefont {M.~E.}\ \bibnamefont {Cates}}, \bibinfo {author} {\bibfnamefont {Y.}~\bibnamefont {Kafri}},\ and\ \bibinfo {author} {\bibfnamefont {J.}~\bibnamefont {Tailleur}},\ }\href {https://doi.org/10.1103/PhysRevE.97.020602} {\bibfield  {journal} {\bibinfo  {journal} {Phys. Rev. E}\ }\textbf {\bibinfo {volume} {97}},\ \bibinfo {pages} {020602(R)} (\bibinfo {year} {2018}{\natexlab{b}})}\BibitemShut {NoStop}%
\bibitem [{\citenamefont {Digregorio}\ \emph {et~al.}(2018)\citenamefont {Digregorio}, \citenamefont {Levis}, \citenamefont {Suma}, \citenamefont {Cugliandolo}, \citenamefont {Gonnella},\ and\ \citenamefont {Pagonabarraga}}]{Digregorio2018FullSeparation}%
  \BibitemOpen
  \bibfield  {author} {\bibinfo {author} {\bibfnamefont {P.}~\bibnamefont {Digregorio}}, \bibinfo {author} {\bibfnamefont {D.}~\bibnamefont {Levis}}, \bibinfo {author} {\bibfnamefont {A.}~\bibnamefont {Suma}}, \bibinfo {author} {\bibfnamefont {L.~F.}\ \bibnamefont {Cugliandolo}}, \bibinfo {author} {\bibfnamefont {G.}~\bibnamefont {Gonnella}},\ and\ \bibinfo {author} {\bibfnamefont {I.}~\bibnamefont {Pagonabarraga}},\ }\href {https://doi.org/10.1103/PhysRevLett.121.098003} {\bibfield  {journal} {\bibinfo  {journal} {Physical Review Letters}\ }\textbf {\bibinfo {volume} {121}},\ \bibinfo {pages} {098003} (\bibinfo {year} {2018})}\BibitemShut {NoStop}%
\bibitem [{\citenamefont {Omar}\ \emph {et~al.}(2021)\citenamefont {Omar}, \citenamefont {Klymko}, \citenamefont {GrandPre},\ and\ \citenamefont {Geissler}}]{Omar2021}%
  \BibitemOpen
  \bibfield  {author} {\bibinfo {author} {\bibfnamefont {A.~K.}\ \bibnamefont {Omar}}, \bibinfo {author} {\bibfnamefont {K.}~\bibnamefont {Klymko}}, \bibinfo {author} {\bibfnamefont {T.}~\bibnamefont {GrandPre}},\ and\ \bibinfo {author} {\bibfnamefont {P.~L.}\ \bibnamefont {Geissler}},\ }\href {https://doi.org/10.1103/PhysRevLett.126.188002} {\bibfield  {journal} {\bibinfo  {journal} {Phys. Rev. Lett.}\ }\textbf {\bibinfo {volume} {126}},\ \bibinfo {pages} {188002} (\bibinfo {year} {2021})}\BibitemShut {NoStop}%
\bibitem [{\citenamefont {Omar}\ \emph {et~al.}(2023)\citenamefont {Omar}, \citenamefont {Row}, \citenamefont {Mallory},\ and\ \citenamefont {Brady}}]{Omar2023b}%
  \BibitemOpen
  \bibfield  {author} {\bibinfo {author} {\bibfnamefont {A.~K.}\ \bibnamefont {Omar}}, \bibinfo {author} {\bibfnamefont {H.}~\bibnamefont {Row}}, \bibinfo {author} {\bibfnamefont {S.~A.}\ \bibnamefont {Mallory}},\ and\ \bibinfo {author} {\bibfnamefont {J.~F.}\ \bibnamefont {Brady}},\ }\href {https://doi.org/10.1073/pnas.2219900120} {\bibfield  {journal} {\bibinfo  {journal} {Proc. Natl. Acad. Sci. U.S.A.}\ }\textbf {\bibinfo {volume} {120}},\ \bibinfo {pages} {e2219900120} (\bibinfo {year} {2023})}\BibitemShut {NoStop}%
\bibitem [{\citenamefont {Patch}\ \emph {et~al.}(2018)\citenamefont {Patch}, \citenamefont {Sussman}, \citenamefont {Yllanes},\ and\ \citenamefont {Marchetti}}]{Patch2018}%
  \BibitemOpen
  \bibfield  {author} {\bibinfo {author} {\bibfnamefont {A.}~\bibnamefont {Patch}}, \bibinfo {author} {\bibfnamefont {D.~M.}\ \bibnamefont {Sussman}}, \bibinfo {author} {\bibfnamefont {D.}~\bibnamefont {Yllanes}},\ and\ \bibinfo {author} {\bibfnamefont {M.~C.}\ \bibnamefont {Marchetti}},\ }\href {https://doi.org/10.1039/C8SM00899J} {\bibfield  {journal} {\bibinfo  {journal} {Soft Matter}\ }\textbf {\bibinfo {volume} {14}},\ \bibinfo {pages} {7435} (\bibinfo {year} {2018})}\BibitemShut {NoStop}%
\bibitem [{\citenamefont {Fausti}\ \emph {et~al.}(2021)\citenamefont {Fausti}, \citenamefont {Tjhung}, \citenamefont {Cates},\ and\ \citenamefont {Nardini}}]{Fausti2021CapillarySeparation}%
  \BibitemOpen
  \bibfield  {author} {\bibinfo {author} {\bibfnamefont {G.}~\bibnamefont {Fausti}}, \bibinfo {author} {\bibfnamefont {E.}~\bibnamefont {Tjhung}}, \bibinfo {author} {\bibfnamefont {M.~E.}\ \bibnamefont {Cates}},\ and\ \bibinfo {author} {\bibfnamefont {C.}~\bibnamefont {Nardini}},\ }\href {https://doi.org/10.1103/PhysRevLett.127.068001} {\bibfield  {journal} {\bibinfo  {journal} {Phys. Rev. Lett.}\ }\textbf {\bibinfo {volume} {127}},\ \bibinfo {pages} {068001} (\bibinfo {year} {2021})}\BibitemShut {NoStop}%
\bibitem [{\citenamefont {Langford}\ and\ \citenamefont {Omar}(2024)}]{Langford2024TheoryPhases}%
  \BibitemOpen
  \bibfield  {author} {\bibinfo {author} {\bibfnamefont {L.}~\bibnamefont {Langford}}\ and\ \bibinfo {author} {\bibfnamefont {A.~K.}\ \bibnamefont {Omar}},\ }\href {https://doi.org/10.1103/PhysRevE.110.054604} {\bibfield  {journal} {\bibinfo  {journal} {Phys. Rev. E}\ }\textbf {\bibinfo {volume} {110}},\ \bibinfo {pages} {054604} (\bibinfo {year} {2024})}\BibitemShut {NoStop}%
\bibitem [{\citenamefont {Redner}\ \emph {et~al.}(2016)\citenamefont {Redner}, \citenamefont {Wagner}, \citenamefont {Baskaran},\ and\ \citenamefont {Hagan}}]{Redner2016}%
  \BibitemOpen
  \bibfield  {author} {\bibinfo {author} {\bibfnamefont {G.~S.}\ \bibnamefont {Redner}}, \bibinfo {author} {\bibfnamefont {C.~G.}\ \bibnamefont {Wagner}}, \bibinfo {author} {\bibfnamefont {A.}~\bibnamefont {Baskaran}},\ and\ \bibinfo {author} {\bibfnamefont {M.~F.}\ \bibnamefont {Hagan}},\ }\href {https://doi.org/10.1103/PhysRevLett.117.148002} {\bibfield  {journal} {\bibinfo  {journal} {Phys. Rev. Lett.}\ }\textbf {\bibinfo {volume} {117}},\ \bibinfo {pages} {148002} (\bibinfo {year} {2016})}\BibitemShut {NoStop}%
\bibitem [{\citenamefont {Cates}\ and\ \citenamefont {Nardini}(2023)}]{Cates2023ClassicalSeparation}%
  \BibitemOpen
  \bibfield  {author} {\bibinfo {author} {\bibfnamefont {M.}~\bibnamefont {Cates}}\ and\ \bibinfo {author} {\bibfnamefont {C.}~\bibnamefont {Nardini}},\ }\href {https://doi.org/10.1103/PhysRevLett.130.098203} {\bibfield  {journal} {\bibinfo  {journal} {Phys. Rev. Lett.}\ }\textbf {\bibinfo {volume} {130}},\ \bibinfo {pages} {098203} (\bibinfo {year} {2023})}\BibitemShut {NoStop}%
\bibitem [{\citenamefont {Langford}\ and\ \citenamefont {Omar}(2025)}]{Langford2025TheMatter}%
  \BibitemOpen
  \bibfield  {author} {\bibinfo {author} {\bibfnamefont {L.}~\bibnamefont {Langford}}\ and\ \bibinfo {author} {\bibfnamefont {A.~K.}\ \bibnamefont {Omar}},\ }\href {https://doi.org/10.1063/5.0263060} {\bibfield  {journal} {\bibinfo  {journal} {The Journal of Chemical Physics}\ }\textbf {\bibinfo {volume} {163}} (\bibinfo {year} {2025})}\BibitemShut {NoStop}%
\bibitem [{\citenamefont {Evans}\ and\ \citenamefont {Omar}(2026)}]{Evans2026TheorySpheres}%
  \BibitemOpen
  \bibfield  {author} {\bibinfo {author} {\bibfnamefont {D.}~\bibnamefont {Evans}}\ and\ \bibinfo {author} {\bibfnamefont {A.~K.}\ \bibnamefont {Omar}},\ }\href {https://doi.org/10.1039/D5SM01196E} {\bibfield  {journal} {\bibinfo  {journal} {Soft Matter}\ }\textbf {\bibinfo {volume} {22}},\ \bibinfo {pages} {1962} (\bibinfo {year} {2026})}\BibitemShut {NoStop}%
\bibitem [{\citenamefont {Bialk{\'{e}}}\ \emph {et~al.}(2015)\citenamefont {Bialk{\'{e}}}, \citenamefont {Siebert}, \citenamefont {L{\"{o}}wen},\ and\ \citenamefont {Speck}}]{Bialke2015}%
  \BibitemOpen
  \bibfield  {author} {\bibinfo {author} {\bibfnamefont {J.}~\bibnamefont {Bialk{\'{e}}}}, \bibinfo {author} {\bibfnamefont {J.~T.}\ \bibnamefont {Siebert}}, \bibinfo {author} {\bibfnamefont {H.}~\bibnamefont {L{\"{o}}wen}},\ and\ \bibinfo {author} {\bibfnamefont {T.}~\bibnamefont {Speck}},\ }\href {https://doi.org/10.1103/PhysRevLett.115.098301} {\bibfield  {journal} {\bibinfo  {journal} {Phys. Rev. Lett.}\ }\textbf {\bibinfo {volume} {115}},\ \bibinfo {pages} {98301} (\bibinfo {year} {2015})}\BibitemShut {NoStop}%
\bibitem [{\citenamefont {Caporusso}\ \emph {et~al.}(2020)\citenamefont {Caporusso}, \citenamefont {Digregorio}, \citenamefont {Levis}, \citenamefont {Cugliandolo},\ and\ \citenamefont {Gonnella}}]{Caporusso2020Motility-InducedSystem}%
  \BibitemOpen
  \bibfield  {author} {\bibinfo {author} {\bibfnamefont {C.~B.}\ \bibnamefont {Caporusso}}, \bibinfo {author} {\bibfnamefont {P.}~\bibnamefont {Digregorio}}, \bibinfo {author} {\bibfnamefont {D.}~\bibnamefont {Levis}}, \bibinfo {author} {\bibfnamefont {L.~F.}\ \bibnamefont {Cugliandolo}},\ and\ \bibinfo {author} {\bibfnamefont {G.}~\bibnamefont {Gonnella}},\ }\href {https://doi.org/10.1103/PhysRevLett.125.178004} {\bibfield  {journal} {\bibinfo  {journal} {Phys. Rev. Lett.}\ }\textbf {\bibinfo {volume} {125}},\ \bibinfo {pages} {178004} (\bibinfo {year} {2020})}\BibitemShut {NoStop}%
\bibitem [{\citenamefont {Nakano}\ and\ \citenamefont {Adachi}(2024)}]{Nakano2024UniversalParticles}%
  \BibitemOpen
  \bibfield  {author} {\bibinfo {author} {\bibfnamefont {H.}~\bibnamefont {Nakano}}\ and\ \bibinfo {author} {\bibfnamefont {K.}~\bibnamefont {Adachi}},\ }\href {https://doi.org/10.1103/PhysRevResearch.6.013074} {\bibfield  {journal} {\bibinfo  {journal} {Physical Review Research}\ }\textbf {\bibinfo {volume} {6}},\ \bibinfo {pages} {013074} (\bibinfo {year} {2024})}\BibitemShut {NoStop}%
\bibitem [{\citenamefont {Cates}\ and\ \citenamefont {Nardini}(2025)}]{Cates2025ActivePhysics}%
  \BibitemOpen
  \bibfield  {author} {\bibinfo {author} {\bibfnamefont {M.~E.}\ \bibnamefont {Cates}}\ and\ \bibinfo {author} {\bibfnamefont {C.}~\bibnamefont {Nardini}},\ }\href {https://doi.org/10.1088/1361-6633/add278} {\bibfield  {journal} {\bibinfo  {journal} {Reports on Progress in Physics}\ }\textbf {\bibinfo {volume} {88}},\ \bibinfo {pages} {056601} (\bibinfo {year} {2025})}\BibitemShut {NoStop}%
\bibitem [{\citenamefont {Stukowski}(2010)}]{Stukowski2010VisualizationTool}%
  \BibitemOpen
  \bibfield  {author} {\bibinfo {author} {\bibfnamefont {A.}~\bibnamefont {Stukowski}},\ }\href {https://doi.org/10.1088/0965-0393/18/1/015012} {\bibfield  {journal} {\bibinfo  {journal} {Model. Simul. Mat. Sci. Eng.}\ }\textbf {\bibinfo {volume} {18}},\ \bibinfo {pages} {015012} (\bibinfo {year} {2010})}\BibitemShut {NoStop}%
\bibitem [{\citenamefont {Fausti}\ \emph {et~al.}(2024)\citenamefont {Fausti}, \citenamefont {Cates},\ and\ \citenamefont {Nardini}}]{Fausti2024StatisticalFluids}%
  \BibitemOpen
  \bibfield  {author} {\bibinfo {author} {\bibfnamefont {G.}~\bibnamefont {Fausti}}, \bibinfo {author} {\bibfnamefont {M.~E.}\ \bibnamefont {Cates}},\ and\ \bibinfo {author} {\bibfnamefont {C.}~\bibnamefont {Nardini}},\ }\href {https://doi.org/10.1103/PhysRevE.110.L042103} {\bibfield  {journal} {\bibinfo  {journal} {Physical Review E}\ }\textbf {\bibinfo {volume} {110}},\ \bibinfo {pages} {L042103} (\bibinfo {year} {2024})}\BibitemShut {NoStop}%
\bibitem [{\citenamefont {Shi}\ \emph {et~al.}(2020)\citenamefont {Shi}, \citenamefont {Fausti}, \citenamefont {Chat{\'{e}}}, \citenamefont {Nardini},\ and\ \citenamefont {Solon}}]{Shi2020Self-OrganizedParticles}%
  \BibitemOpen
  \bibfield  {author} {\bibinfo {author} {\bibfnamefont {X.-q.}\ \bibnamefont {Shi}}, \bibinfo {author} {\bibfnamefont {G.}~\bibnamefont {Fausti}}, \bibinfo {author} {\bibfnamefont {H.}~\bibnamefont {Chat{\'{e}}}}, \bibinfo {author} {\bibfnamefont {C.}~\bibnamefont {Nardini}},\ and\ \bibinfo {author} {\bibfnamefont {A.}~\bibnamefont {Solon}},\ }\href {https://doi.org/10.1103/PhysRevLett.125.168001} {\bibfield  {journal} {\bibinfo  {journal} {Phys. Rev. Lett.}\ }\textbf {\bibinfo {volume} {125}},\ \bibinfo {pages} {168001} (\bibinfo {year} {2020})}\BibitemShut {NoStop}%
\bibitem [{\citenamefont {Yan}\ \emph {et~al.}(2025)\citenamefont {Yan}, \citenamefont {Frey}, \citenamefont {M{\"{u}}ller},\ and\ \citenamefont {Klumpp}}]{Yan2025StochasticMatter}%
  \BibitemOpen
  \bibfield  {author} {\bibinfo {author} {\bibfnamefont {M.}~\bibnamefont {Yan}}, \bibinfo {author} {\bibfnamefont {E.}~\bibnamefont {Frey}}, \bibinfo {author} {\bibfnamefont {M.}~\bibnamefont {M{\"{u}}ller}},\ and\ \bibinfo {author} {\bibfnamefont {S.}~\bibnamefont {Klumpp}},\ }\href {https://doi.org/10.48550/arXiv.2501.11442} {\bibfield  {journal} {\bibinfo  {journal} {arXiv:2501.11442}\ } (\bibinfo {year} {2025})}\BibitemShut {NoStop}%
\bibitem [{\citenamefont {Weeks}\ \emph {et~al.}(1971)\citenamefont {Weeks}, \citenamefont {Chandler},\ and\ \citenamefont {Andersen}}]{Weeks1971}%
  \BibitemOpen
  \bibfield  {author} {\bibinfo {author} {\bibfnamefont {J.~D.}\ \bibnamefont {Weeks}}, \bibinfo {author} {\bibfnamefont {D.}~\bibnamefont {Chandler}},\ and\ \bibinfo {author} {\bibfnamefont {H.~C.}\ \bibnamefont {Andersen}},\ }\href {https://doi.org/10.1063/1.1674820} {\bibfield  {journal} {\bibinfo  {journal} {J. Chem. Phys.}\ }\textbf {\bibinfo {volume} {54}},\ \bibinfo {pages} {5237} (\bibinfo {year} {1971})}\BibitemShut {NoStop}%
\bibitem [{\citenamefont {Omar}\ \emph {et~al.}(2020)\citenamefont {Omar}, \citenamefont {Wang},\ and\ \citenamefont {Brady}}]{Omar2020}%
  \BibitemOpen
  \bibfield  {author} {\bibinfo {author} {\bibfnamefont {A.~K.}\ \bibnamefont {Omar}}, \bibinfo {author} {\bibfnamefont {Z.-G.}\ \bibnamefont {Wang}},\ and\ \bibinfo {author} {\bibfnamefont {J.~F.}\ \bibnamefont {Brady}},\ }\href {https://doi.org/10.1103/PhysRevE.101.012604} {\bibfield  {journal} {\bibinfo  {journal} {Phys. Rev. E}\ }\textbf {\bibinfo {volume} {101}},\ \bibinfo {pages} {012604} (\bibinfo {year} {2020})}\BibitemShut {NoStop}%
\bibitem [{\citenamefont {Mallory}\ \emph {et~al.}(2021)\citenamefont {Mallory}, \citenamefont {Omar},\ and\ \citenamefont {Brady}}]{Mallory2021}%
  \BibitemOpen
  \bibfield  {author} {\bibinfo {author} {\bibfnamefont {S.~A.}\ \bibnamefont {Mallory}}, \bibinfo {author} {\bibfnamefont {A.~K.}\ \bibnamefont {Omar}},\ and\ \bibinfo {author} {\bibfnamefont {J.~F.}\ \bibnamefont {Brady}},\ }\href {https://doi.org/10.1103/PhysRevE.104.044612} {\bibfield  {journal} {\bibinfo  {journal} {Phys. Rev. E}\ }\textbf {\bibinfo {volume} {104}},\ \bibinfo {pages} {044612} (\bibinfo {year} {2021})}\BibitemShut {NoStop}%
\bibitem [{\citenamefont {Redner}\ \emph {et~al.}(2013{\natexlab{b}})\citenamefont {Redner}, \citenamefont {Baskaran},\ and\ \citenamefont {Hagan}}]{Redner2013ReentrantAttraction}%
  \BibitemOpen
  \bibfield  {author} {\bibinfo {author} {\bibfnamefont {G.~S.}\ \bibnamefont {Redner}}, \bibinfo {author} {\bibfnamefont {A.}~\bibnamefont {Baskaran}},\ and\ \bibinfo {author} {\bibfnamefont {M.~F.}\ \bibnamefont {Hagan}},\ }\href {https://doi.org/10.1103/PhysRevE.88.012305} {\bibfield  {journal} {\bibinfo  {journal} {Physical Review E}\ }\textbf {\bibinfo {volume} {88}},\ \bibinfo {pages} {012305} (\bibinfo {year} {2013}{\natexlab{b}})}\BibitemShut {NoStop}%
\bibitem [{\citenamefont {Martin-Roca}\ \emph {et~al.}(2021)\citenamefont {Martin-Roca}, \citenamefont {Martinez}, \citenamefont {Alexander}, \citenamefont {Diez}, \citenamefont {Aarts}, \citenamefont {Alarcon}, \citenamefont {Ram{\'{i}}rez},\ and\ \citenamefont {Valeriani}}]{Martin-Roca2021CharacterizationFeatures}%
  \BibitemOpen
  \bibfield  {author} {\bibinfo {author} {\bibfnamefont {J.}~\bibnamefont {Martin-Roca}}, \bibinfo {author} {\bibfnamefont {R.}~\bibnamefont {Martinez}}, \bibinfo {author} {\bibfnamefont {L.~C.}\ \bibnamefont {Alexander}}, \bibinfo {author} {\bibfnamefont {A.~L.}\ \bibnamefont {Diez}}, \bibinfo {author} {\bibfnamefont {D.~G. A.~L.}\ \bibnamefont {Aarts}}, \bibinfo {author} {\bibfnamefont {F.}~\bibnamefont {Alarcon}}, \bibinfo {author} {\bibfnamefont {J.}~\bibnamefont {Ram{\'{i}}rez}},\ and\ \bibinfo {author} {\bibfnamefont {C.}~\bibnamefont {Valeriani}},\ }\href {https://doi.org/10.1063/5.0040141} {\bibfield  {journal} {\bibinfo  {journal} {The Journal of Chemical Physics}\ }\textbf {\bibinfo {volume} {154}} (\bibinfo {year} {2021})}\BibitemShut {NoStop}%
\bibitem [{\citenamefont {Z{\"{o}}ttl}\ and\ \citenamefont {Stark}(2023)}]{Zottl2023ModelingFields}%
  \BibitemOpen
  \bibfield  {author} {\bibinfo {author} {\bibfnamefont {A.}~\bibnamefont {Z{\"{o}}ttl}}\ and\ \bibinfo {author} {\bibfnamefont {H.}~\bibnamefont {Stark}},\ }\href {https://doi.org/10.1146/annurev-conmatphys-040821-115500} {\bibfield  {journal} {\bibinfo  {journal} {Annual Review of Condensed Matter Physics}\ }\textbf {\bibinfo {volume} {14}},\ \bibinfo {pages} {109} (\bibinfo {year} {2023})}\BibitemShut {NoStop}%
\bibitem [{Note1()}]{Note1}%
  \BibitemOpen
  \bibinfo {note} {We note that the particle activity can also be quantified by defining a P\'{e}clet number $\protect \text {Pe}\equiv U_o\sigma /D_T$. We can express this P\'{e}clet via our dimensionless run length and thermalization coefficient: $\protect \text {Pe} = \ell _o/\left (\protect \mathcal {T}\sigma \right )$.}\BibitemShut {Stop}%
\bibitem [{Note2()}]{Note2}%
  \BibitemOpen
  \bibinfo {note} {The significance of this difference in overall area fractions, as well as simulation results for a thermal polydisperse simulation done at $\phi _o=0.64$, is further discussed in the SM.}\BibitemShut {Stop}%
\bibitem [{\citenamefont {Anderson}\ \emph {et~al.}(2020)\citenamefont {Anderson}, \citenamefont {Glaser},\ and\ \citenamefont {Glotzer}}]{Anderson2020}%
  \BibitemOpen
  \bibfield  {author} {\bibinfo {author} {\bibfnamefont {J.~A.}\ \bibnamefont {Anderson}}, \bibinfo {author} {\bibfnamefont {J.}~\bibnamefont {Glaser}},\ and\ \bibinfo {author} {\bibfnamefont {S.~C.}\ \bibnamefont {Glotzer}},\ }\href {https://doi.org/10.1016/j.commatsci.2019.109363} {\bibfield  {journal} {\bibinfo  {journal} {Comput. Mater. Sci.}\ }\textbf {\bibinfo {volume} {173}},\ \bibinfo {pages} {109363} (\bibinfo {year} {2020})}\BibitemShut {NoStop}%
\bibitem [{Note3()}]{Note3}%
  \BibitemOpen
  \bibinfo {note} {$n(A)$ is thus generally an extensive quantity that can be made intensive by, for example, dividing by the area of the liquid phase~\cite {Shi2020Self-OrganizedParticles}.}\BibitemShut {Stop}%
\bibitem [{\citenamefont {Ester}\ \emph {et~al.}(1996)\citenamefont {Ester}, \citenamefont {Kriegel}, \citenamefont {Sander},\ and\ \citenamefont {Xu}}]{Ester1996ANoise}%
  \BibitemOpen
  \bibfield  {author} {\bibinfo {author} {\bibfnamefont {M.}~\bibnamefont {Ester}}, \bibinfo {author} {\bibfnamefont {H.-P.}\ \bibnamefont {Kriegel}}, \bibinfo {author} {\bibfnamefont {J.}~\bibnamefont {Sander}},\ and\ \bibinfo {author} {\bibfnamefont {X.}~\bibnamefont {Xu}},\ }in\ \href {https://dl.acm.org/doi/10.5555/3001460.3001507} {\emph {\bibinfo {booktitle} {Proceedings of the Second International Conference on Knowledge Discovery and Data Mining}}}\ (\bibinfo  {publisher} {AAAI Press},\ \bibinfo {year} {1996})\BibitemShut {NoStop}%
\bibitem [{Note4()}]{Note4}%
  \BibitemOpen
  \bibinfo {note} {See Supplemental Material at [URL]}\BibitemShut {NoStop}%
\bibitem [{\citenamefont {Caprini}\ \emph {et~al.}(2020{\natexlab{a}})\citenamefont {Caprini}, \citenamefont {Marconi}, \citenamefont {Maggi}, \citenamefont {Paoluzzi},\ and\ \citenamefont {Puglisi}}]{Caprini2020HiddenDisks}%
  \BibitemOpen
  \bibfield  {author} {\bibinfo {author} {\bibfnamefont {L.}~\bibnamefont {Caprini}}, \bibinfo {author} {\bibfnamefont {U.~M.~B.}\ \bibnamefont {Marconi}}, \bibinfo {author} {\bibfnamefont {C.}~\bibnamefont {Maggi}}, \bibinfo {author} {\bibfnamefont {M.}~\bibnamefont {Paoluzzi}},\ and\ \bibinfo {author} {\bibfnamefont {A.}~\bibnamefont {Puglisi}},\ }\href {https://doi.org/10.1103/PhysRevResearch.2.023321 Export Citation} {\bibfield  {journal} {\bibinfo  {journal} {Physical Review Research}\ }\textbf {\bibinfo {volume} {2}},\ \bibinfo {pages} {023321} (\bibinfo {year} {2020}{\natexlab{a}})}\BibitemShut {NoStop}%
\bibitem [{\citenamefont {Caprini}\ \emph {et~al.}(2020{\natexlab{b}})\citenamefont {Caprini}, \citenamefont {Marini Bettolo~Marconi},\ and\ \citenamefont {Puglisi}}]{Caprini2020SpontaneousSeparation}%
  \BibitemOpen
  \bibfield  {author} {\bibinfo {author} {\bibfnamefont {L.}~\bibnamefont {Caprini}}, \bibinfo {author} {\bibfnamefont {U.}~\bibnamefont {Marini Bettolo~Marconi}},\ and\ \bibinfo {author} {\bibfnamefont {A.}~\bibnamefont {Puglisi}},\ }\href {https://doi.org/10.1103/PhysRevLett.124.078001} {\bibfield  {journal} {\bibinfo  {journal} {Physical Review Letters}\ }\textbf {\bibinfo {volume} {124}},\ \bibinfo {pages} {078001} (\bibinfo {year} {2020}{\natexlab{b}})}\BibitemShut {NoStop}%
\bibitem [{\citenamefont {Caprini}\ and\ \citenamefont {Marini Bettolo~Marconi}(2021)}]{Caprini2021SpatialParticles}%
  \BibitemOpen
  \bibfield  {author} {\bibinfo {author} {\bibfnamefont {L.}~\bibnamefont {Caprini}}\ and\ \bibinfo {author} {\bibfnamefont {U.}~\bibnamefont {Marini Bettolo~Marconi}},\ }\href {https://doi.org/10.1039/D0SM02273J} {\bibfield  {journal} {\bibinfo  {journal} {Soft Matter}\ }\textbf {\bibinfo {volume} {17}},\ \bibinfo {pages} {4109} (\bibinfo {year} {2021})}\BibitemShut {NoStop}%
\bibitem [{\citenamefont {Takatori}\ \emph {et~al.}(2014)\citenamefont {Takatori}, \citenamefont {Yan},\ and\ \citenamefont {Brady}}]{Takatori2014}%
  \BibitemOpen
  \bibfield  {author} {\bibinfo {author} {\bibfnamefont {S.~C.}\ \bibnamefont {Takatori}}, \bibinfo {author} {\bibfnamefont {W.}~\bibnamefont {Yan}},\ and\ \bibinfo {author} {\bibfnamefont {J.~F.}\ \bibnamefont {Brady}},\ }\href {https://journals.aps.org/prl/abstract/10.1103/PhysRevLett.113.028103} {\bibfield  {journal} {\bibinfo  {journal} {Phys. Rev. Lett.}\ }\textbf {\bibinfo {volume} {113}},\ \bibinfo {pages} {028103} (\bibinfo {year} {2014})}\BibitemShut {NoStop}%
\bibitem [{\citenamefont {Vrugt}\ \emph {et~al.}(2023)\citenamefont {Vrugt}, \citenamefont {Bickmann},\ and\ \citenamefont {Wittkowski}}]{Vrugt2023HowTutorial}%
  \BibitemOpen
  \bibfield  {author} {\bibinfo {author} {\bibfnamefont {M.~t.}\ \bibnamefont {Vrugt}}, \bibinfo {author} {\bibfnamefont {J.}~\bibnamefont {Bickmann}},\ and\ \bibinfo {author} {\bibfnamefont {R.}~\bibnamefont {Wittkowski}},\ }\href {https://doi.org/10.1088/1361-648X/acc440} {\bibfield  {journal} {\bibinfo  {journal} {J. Phys. Condens. Matter}\ }\textbf {\bibinfo {volume} {35}},\ \bibinfo {pages} {313001} (\bibinfo {year} {2023})}\BibitemShut {NoStop}%
\bibitem [{\citenamefont {Phillips}\ \emph {et~al.}(2010)\citenamefont {Phillips}, \citenamefont {Iacovella},\ and\ \citenamefont {Glotzer}}]{Phillips2010StabilitySystems}%
  \BibitemOpen
  \bibfield  {author} {\bibinfo {author} {\bibfnamefont {C.~L.}\ \bibnamefont {Phillips}}, \bibinfo {author} {\bibfnamefont {C.~R.}\ \bibnamefont {Iacovella}},\ and\ \bibinfo {author} {\bibfnamefont {S.~C.}\ \bibnamefont {Glotzer}},\ }\href {https://doi.org/10.1039/B911140A} {\bibfield  {journal} {\bibinfo  {journal} {Soft Matter}\ }\textbf {\bibinfo {volume} {6}},\ \bibinfo {pages} {1693} (\bibinfo {year} {2010})}\BibitemShut {NoStop}%
\bibitem [{\citenamefont {Ramasubramani}\ \emph {et~al.}(2020)\citenamefont {Ramasubramani}, \citenamefont {Dice}, \citenamefont {Harper}, \citenamefont {Spellings}, \citenamefont {Anderson},\ and\ \citenamefont {Glotzer}}]{Ramasubramani2020Freud:Data}%
  \BibitemOpen
  \bibfield  {author} {\bibinfo {author} {\bibfnamefont {V.}~\bibnamefont {Ramasubramani}}, \bibinfo {author} {\bibfnamefont {B.~D.}\ \bibnamefont {Dice}}, \bibinfo {author} {\bibfnamefont {E.~S.}\ \bibnamefont {Harper}}, \bibinfo {author} {\bibfnamefont {M.~P.}\ \bibnamefont {Spellings}}, \bibinfo {author} {\bibfnamefont {J.~A.}\ \bibnamefont {Anderson}},\ and\ \bibinfo {author} {\bibfnamefont {S.~C.}\ \bibnamefont {Glotzer}},\ }\href {https://doi.org/10.1016/j.cpc.2020.107275} {\bibfield  {journal} {\bibinfo  {journal} {Computer Physics Communications}\ }\textbf {\bibinfo {volume} {254}},\ \bibinfo {pages} {107275} (\bibinfo {year} {2020})}\BibitemShut {NoStop}%
\end{thebibliography}
\end{document}


\maketitle

\pagenumbering{roman}

\setcounter{secnumdepth}{2}
\setcounter{tocdepth}{2}
{
	\hypersetup{
		linkcolor=Black,
		citecolor=Black
	}
	\vspace{-30pt}
	\tableofcontents
}

\noindent\rule{5.0cm}{0.4pt}
	{\footnotesize
	$\\$
	{
	    {$^\dag \,$}aomar@berkeley.edu}
	}

\newpage 
\pagenumbering{arabic}
\section{\label{sec:sim}Simulations}
\subsection{\label{sec:critpoint}Impact of Thermalization on Critical Point}
In athermal ABPs, the dynamic pressure controlling the bulk coexistence criteria~\cite{Omar2023b} was determined to be:
\begin{subequations}
    \label{seq:dynpressure}
    \begin{align}
        \mathcal{P} &= p^{\rm act}(\phi) + p^{\rm c}(\phi),
    \end{align}
    \begin{align}
        a_op^{\rm act}(\phi) &= \frac{\zeta U_o \overline{U}\ell_o \phi}{2},
    \end{align}
\end{subequations}
where $\phi$ is the local packing fraction, $\overline{U}$ is a dimensionless effective active speed that is between $0$ and $1$, and $a_o = \pi d^2/4$ is the single particle area. 
Here $p^{\rm c}(\phi)$ is the conservative pressure due to particle collisions.
Here, $d=2^{1/6}\sigma$ corresponds to the hard particle diameter.
The critical point, which represents the onset of the spinodal criteria, can be found by setting both the first and second derivative of $\mathcal{P}(\phi)$ to zero and solving for the critical packing fraction and activity $(\phi_c,\ell_c)$.
When translational noise is introduced, $\mathcal{P}$ is modified in a few ways. 
First, an \textit{ideal gas} pressure becomes present, and is proportional to $k_BT\phi$:
\begin{equation}
    a_op^{\rm ig} = \zeta D_T\phi  = \frac{\mathcal{T}\sigma^2U_o}{\ell_o}\phi,\label{seq:ig}
\end{equation}
Additionally, the equations of state for $\overline{U}$ and $p^{\rm c}(\phi)$ may change.
We will assume that the equations of state do not change and remain of the form used in Refs.~\cite{Takatori2015TowardsMatter,Langford2025PhaseMatter}:
\begin{subequations}
    \label{seq:2deos}
    \begin{align}
        \frac{p^{\rm c}d}{\zeta U_o} &= \frac{8}{\pi}\phi^2\left(1 - \frac{\phi}{\phi_o}\right)^{-1}, 
    \end{align}
    \begin{align}
        \overline{U} &= \left[1 + \phi\left(1-\frac{\phi}{\phi_o}\right)^{-1}\right]^{-1},
    \end{align}
\end{subequations}
where $\phi_o\equiv 0.9$ approximates the close packing fraction for two dimensions.
Using Eqs.~\eqref{seq:dynpressure},~\eqref{seq:ig}, and~\eqref{seq:2deos}, we calculate $\phi_c$ and $\ell_c$ as a function of $\mathcal{T}$.
\begin{figure}
	\centering
	\includegraphics[width=0.9\textwidth]{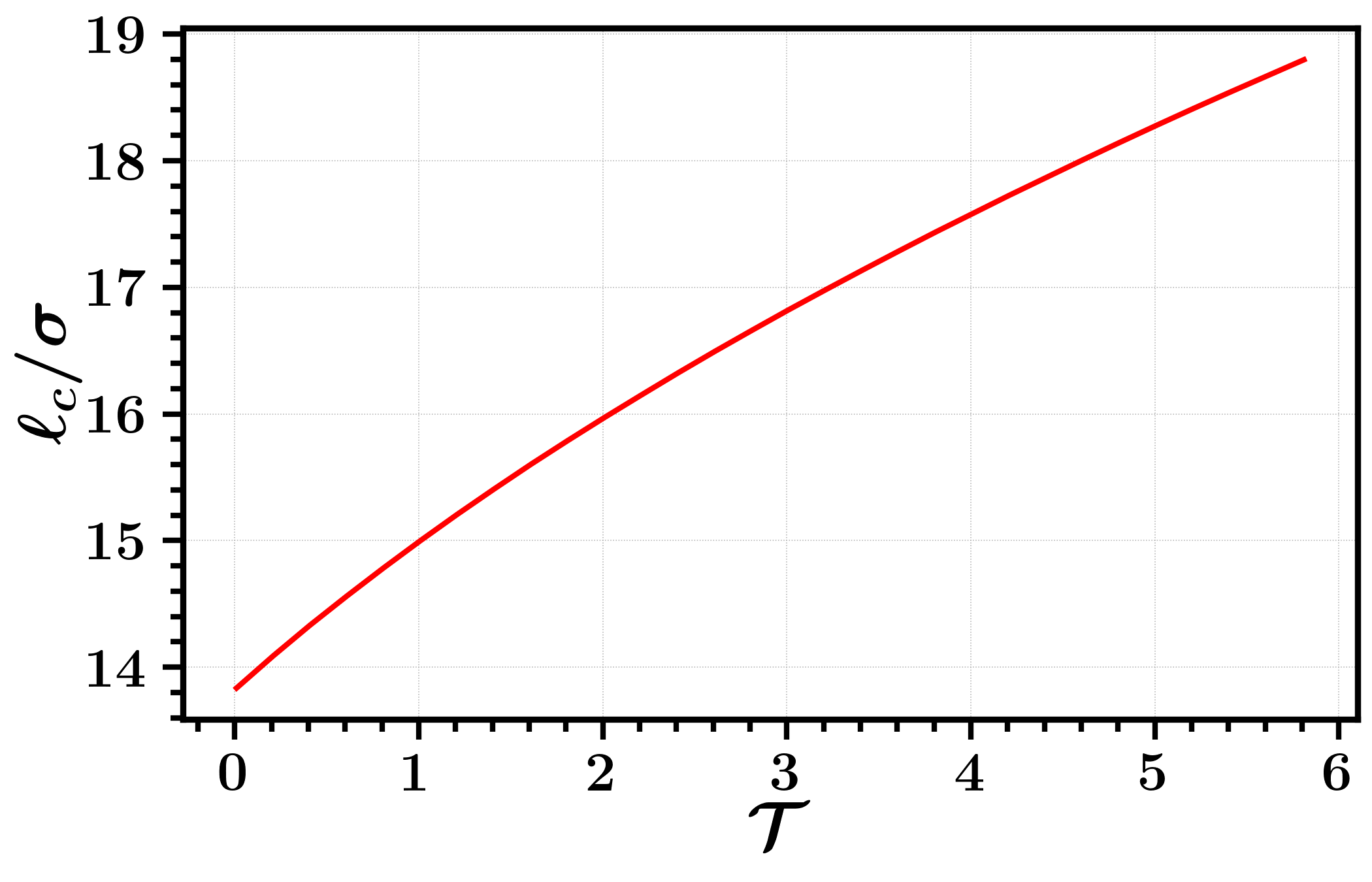}
	\caption{Critical activity $\ell_c$ as a function of thermalization $\mathcal{T}$.}
	\label{fig:critact}
\end{figure}
\begin{figure}
	\centering
	\includegraphics[width=0.9\textwidth]{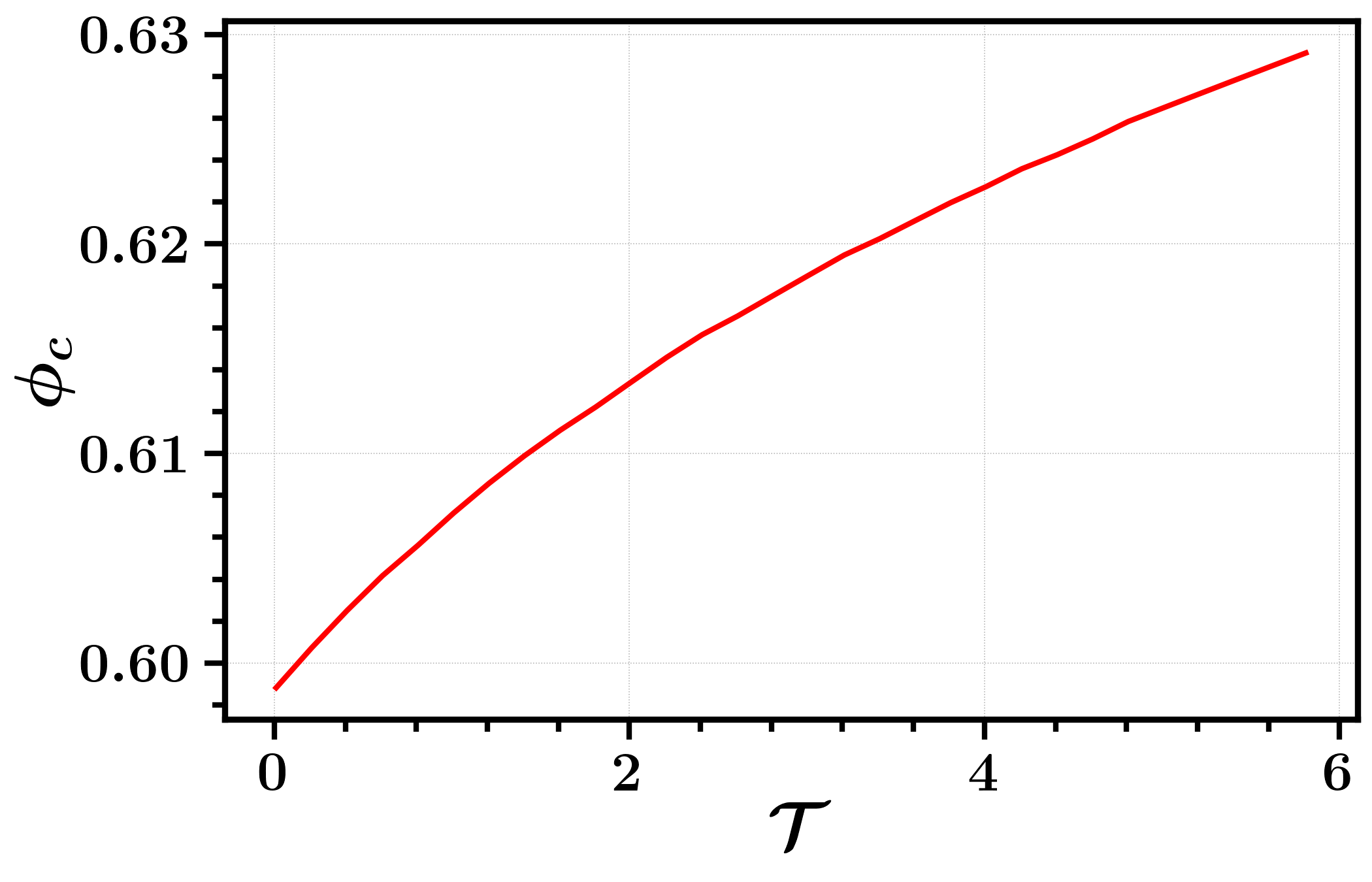}
	\caption{Critical area fraction $\phi_c$ as a function of thermalization $\mathcal{T}$.}
	\label{fig:critdense}
\end{figure}

Based on the results of Figs.~\ref{fig:critact} and~\ref{fig:critdense}, we restrict our thermalization to be less than one $(\mathcal{T}<1)$ in order not to shift the critical activity by more than ten percent. 
\subsection{\label{sec:steadystate}Steady State Verification}
To check that our simulations have reached a steady state, we verify that measured quantities of interest indeed are time independent.
In particular, we can average the measured quantities over smaller chunks of time and verify a lack of systematic drift over time in those measured quantities.
We choose to do this for two quantities that were measured for all simulations: the power-law exponent $\omega$ of $n(A)$, and the bubble area fraction $\phi_B$.
These measurements are broken up into time chunks of $2500~ \sigma/U_o$ to average over, and plotted across the second half of the $100000~U_o/\sigma$ trajectory where statistics reported in the main text were collected.
While the measured quantities could vary significantly, especially for simulations where bubbles were rarefied, no systematic drift was identified.
\begin{figure}
	\centering
	\includegraphics[width=0.9\textwidth]{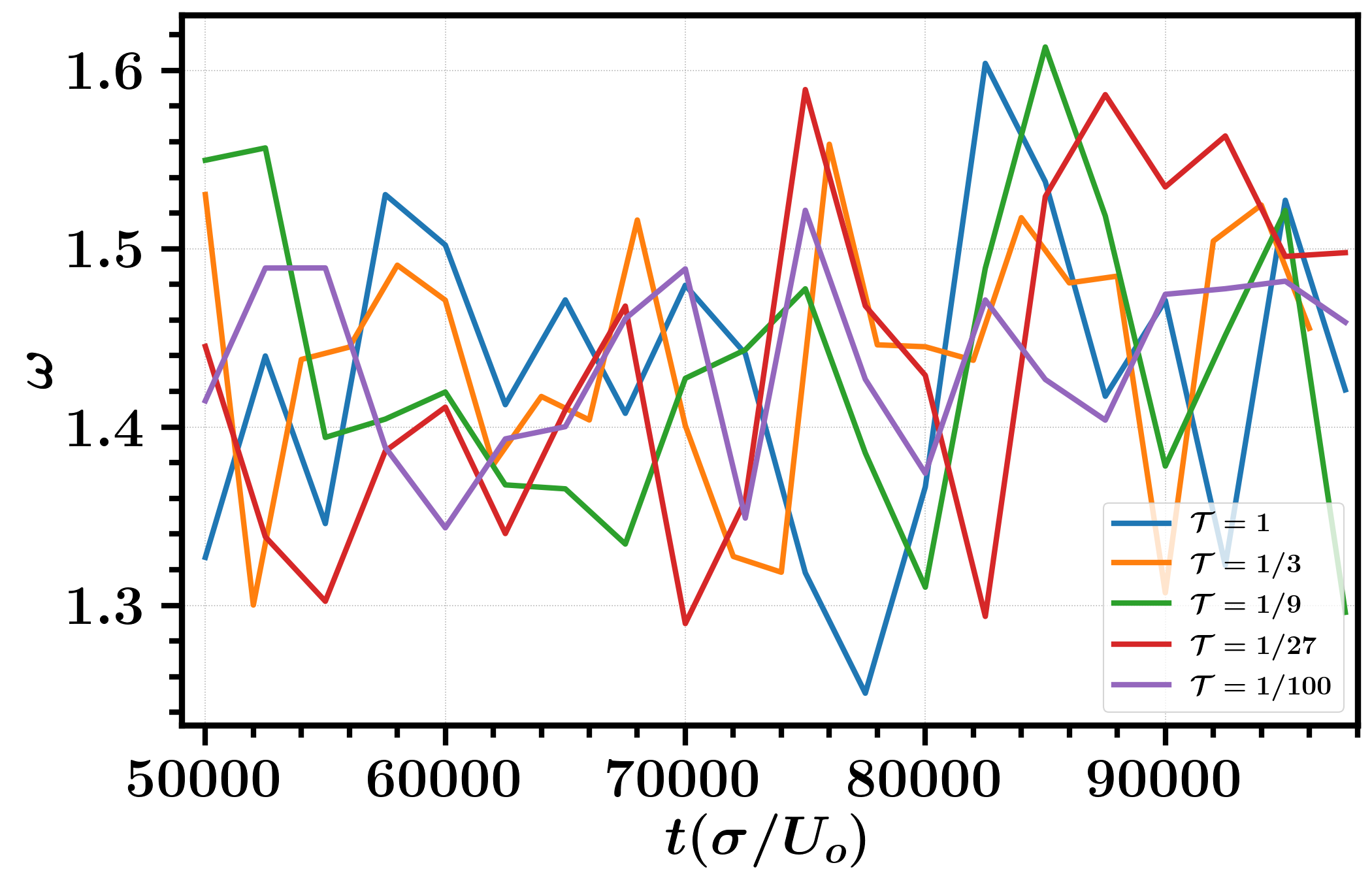}
	\caption{Fits of bubble distribution power law exponent $\omega$ over time for thermalized monodisperse systems.}
	\label{fig:omegamonothermal}
\end{figure}
\begin{figure}
	\centering
	\includegraphics[width=0.9\textwidth]{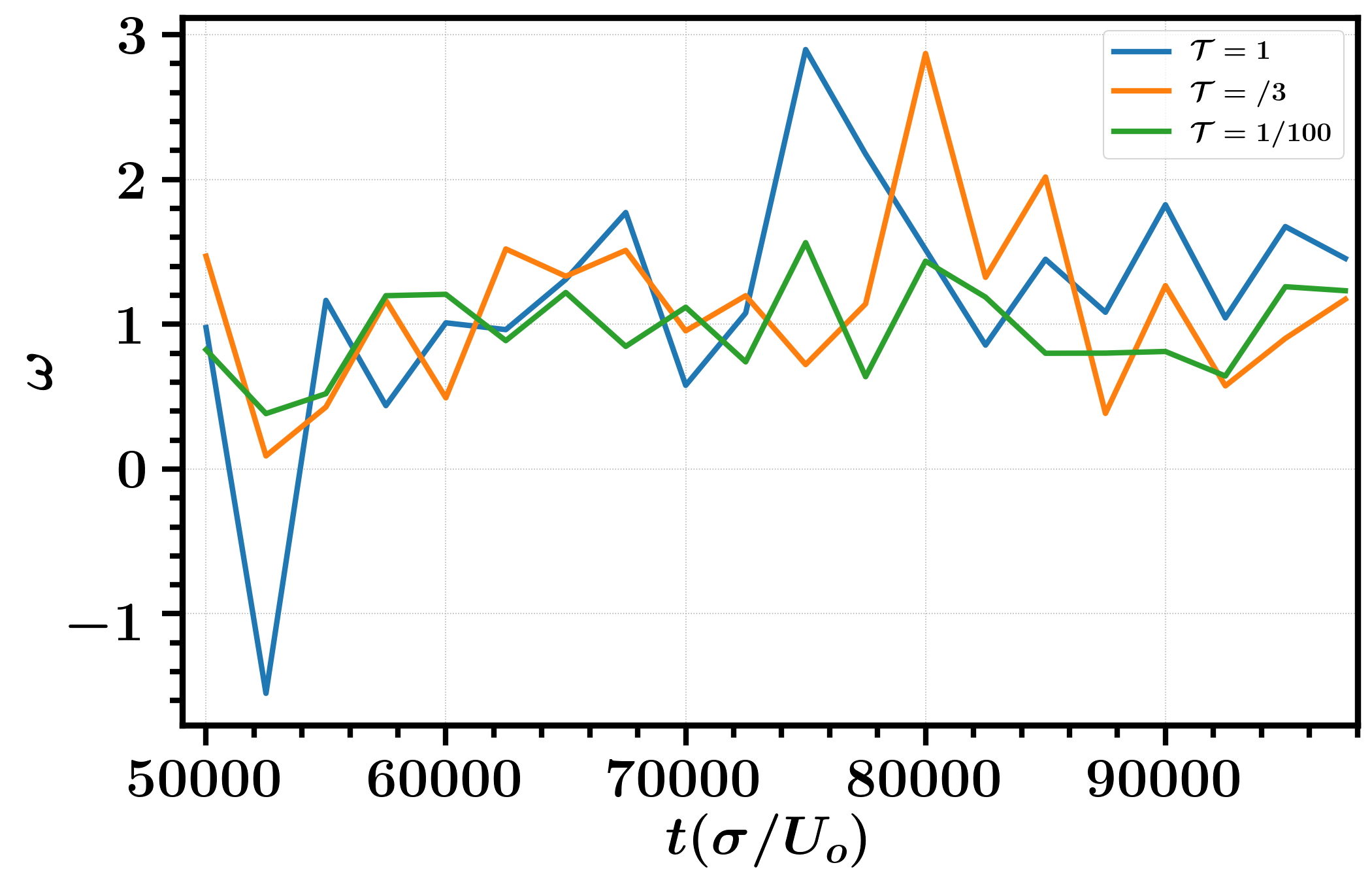}
	\caption{Fits of bubble distribution power law exponent $\omega$ over time for thermalized polydisperse systems.}
	\label{fig:omegapolythermal}
\end{figure}
\begin{figure}
	\centering
	\includegraphics[width=0.9\textwidth]{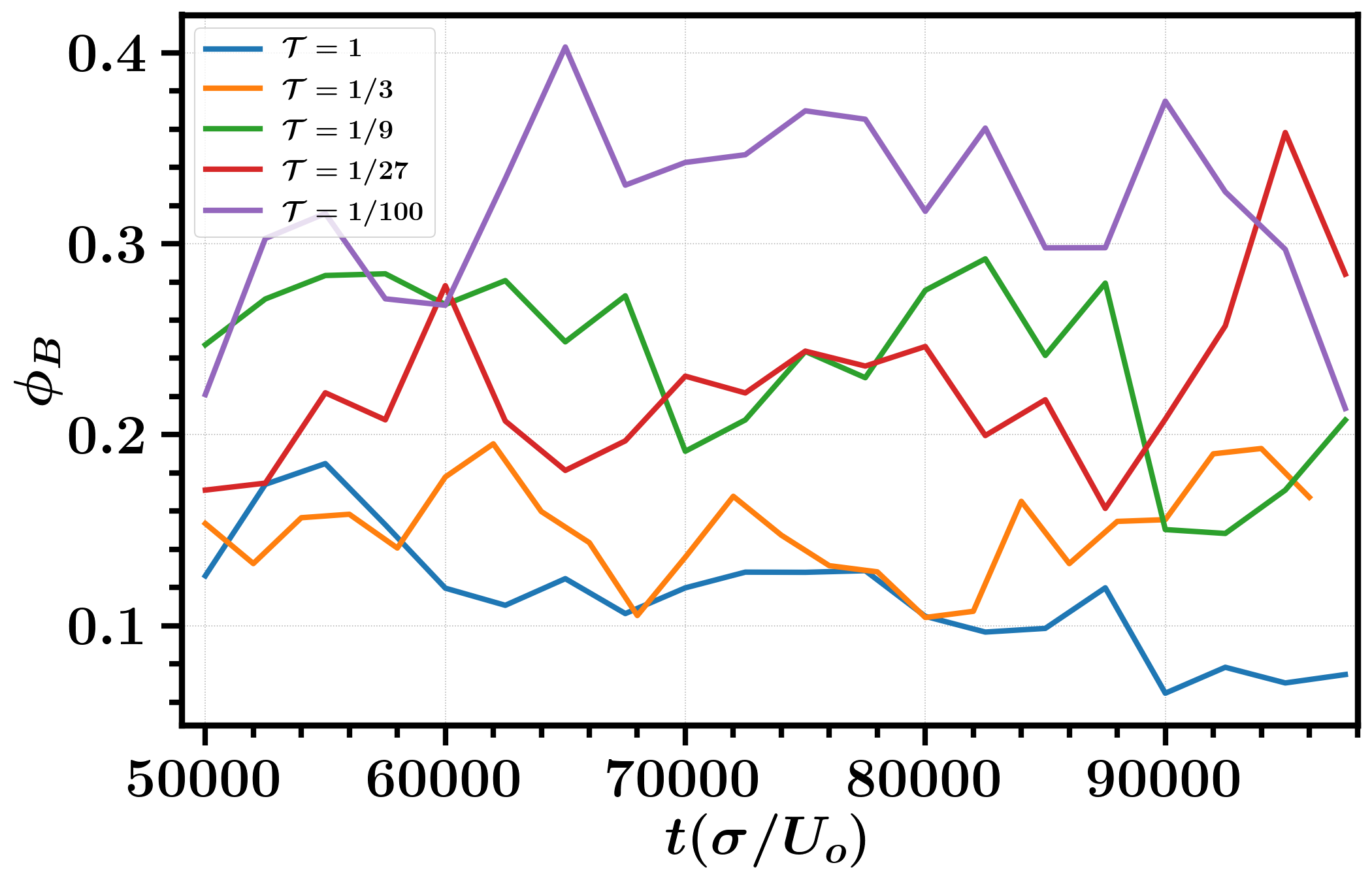}
	\caption{Average bubble fraction $\phi_B$ over time for thermalized monodisperse systems.}
	\label{fig:fracmonothermal}
\end{figure}
\begin{figure}
	\centering
	\includegraphics[width=0.9\textwidth]{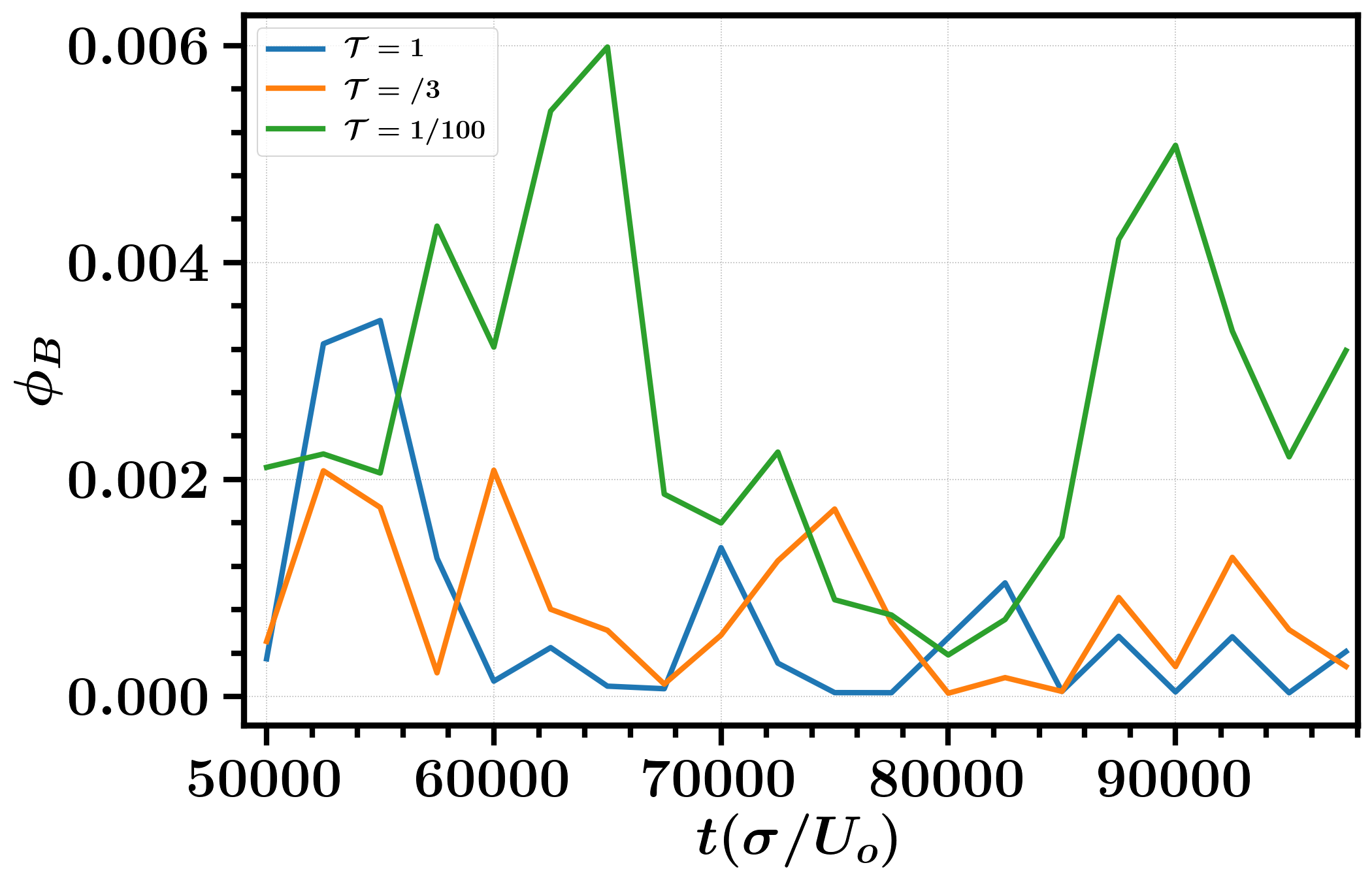}
	\caption{Average bubble fraction $\phi_B$ over time for thermalized polydisperse systems.}
	\label{fig:fracpolythermal}
\end{figure}
\begin{figure}
	\centering
	\includegraphics[width=0.9\textwidth]{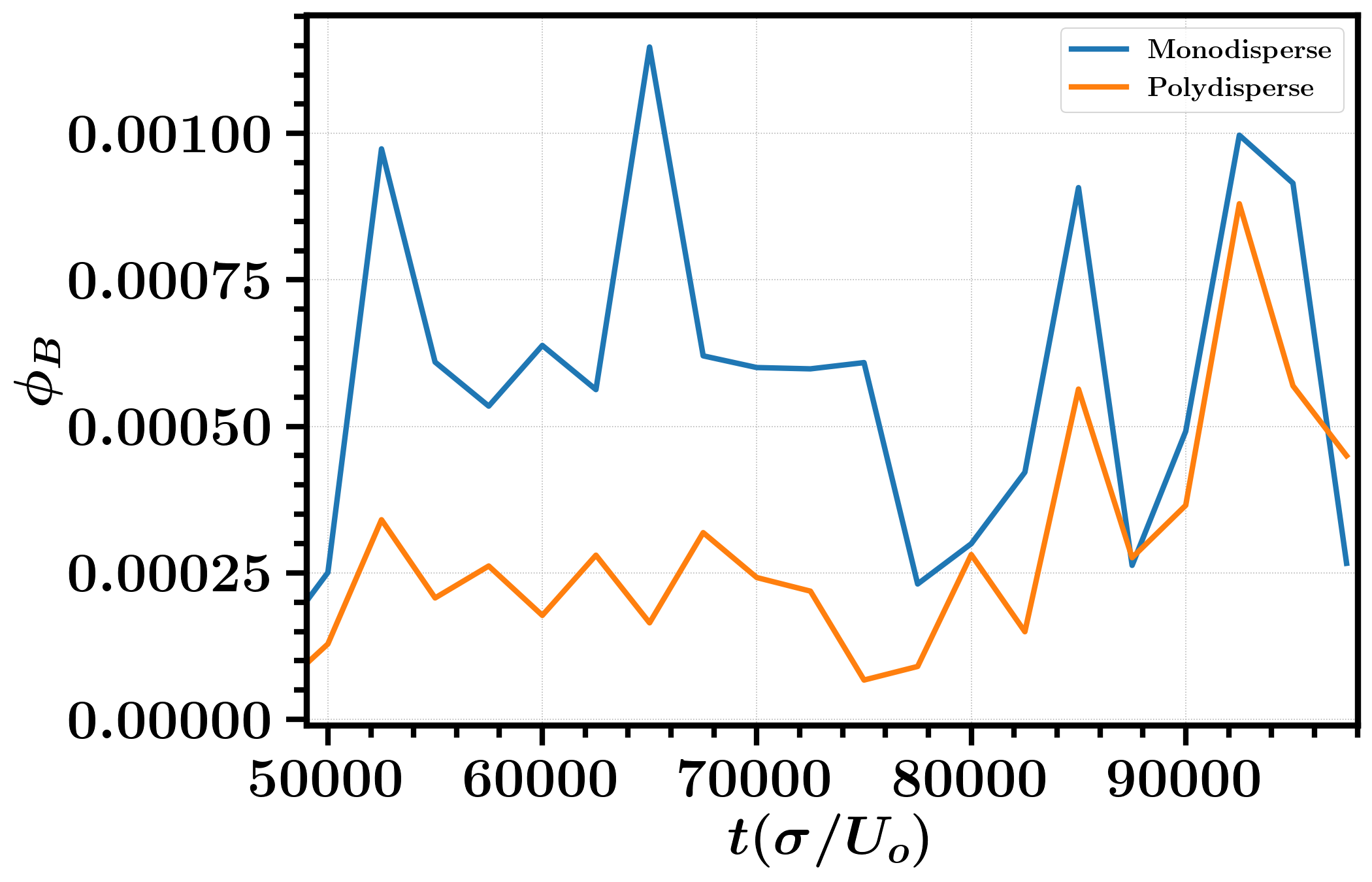}
	\caption{Average bubble fraction $\phi_B$ over time for athermal systems.}
	\label{fig:fracathermal}
\end{figure} \newpage
\section{\label{sec:anly}Analysis}
\subsection{\label{sec:bubble}Bubble Statistics}
In order to record the bubble statistics $n(A)$ and $P(A)$, an algorithm for identifying bubbles based on the method used in Ref.~\cite{Caporusso2020Motility-InducedSystem} was employed.
First, the local coarse-grained area fraction $\tilde{\phi}$ calculated using the coarse-graining function:
\begin{equation}
    \tilde{\phi}(\mathbf{r}) = \sum_{i=1}^{N}\pi \left(\frac{d_i}{2}\right)^2\frac{1}{2\pi \xi^2}\text{exp}\left[-\frac{|\mathbf{r}-\mathbf{r}_i|^2}{2\xi^2}\right],
\end{equation}
where $\xi$ is a coarse-graining length that we set to $2.5\sigma$ for all calculations and $d_i = \sigma_i + (2^{1/6}-1)\sigma$ is the particle diameter.
$\tilde{\phi}(\mathbf{r})$ was calculated on a square grid of sample points with spacing of $3.2\sigma$. 
Grid points with a value of $\tilde{\phi}$ less than some cutoff density $\phi_{\rm cut}$ were labeled as gas phase. 
In all calculations we set $\phi_{\rm cut} = 0.65$
The cluster analysis algorithm DBSCAN~\cite{Ester1996ANoise} was then used to identify clusters of grid points labeled as gas phase.
DBSCAN uses the following rules for determining clusters:
\begin{itemize}
    \item Two objects are considered neighbors if they are within some cutoff distance $\epsilon_{\rm cut}$ of each other. An object is considered a ``core'' object if it has $n_{\rm min}$ neighbors.
    \item Clusters are defined by the connected pathways of all core objects and neighbors of core objects.
    \item Objects that are not core or neighbors of cores are not part of any cluster.
\end{itemize}
Bubbles were identified by using DBSCAN on all points labeled as gas phase with $\epsilon_{\rm cut}$ set to be the grid spacing $3.2\sigma$, and $n_{\rm min} = 4$ corresponding to a square grid.
The largest bubble/cluster identified in each frame was the bulk gas phase, which we ignore.
The area of all other bubbles was calculated, and all bubbles with area larger than $A_{\rm min} = 16\pi \sigma^2$ were added to a histogram with binwidth $\Delta A = 10\sigma^2$.
$n(A)$ was calculated by normalizing this histogram by the number of frames considered and dividing by the bin width. 
$P(A)$ was calculated by normalizing $n(A)$ by the average number of bubbles observed, $N_{\rm bub}$.
Both quantities were then divided by the histogram bin spacing in order to maintain the correct units.
$\phi_B$ was calculated by recording the total area of identified bubbles and dividing by the total area of gas phase.

\begin{figure}
	\centering
	\includegraphics[width=0.9\textwidth]{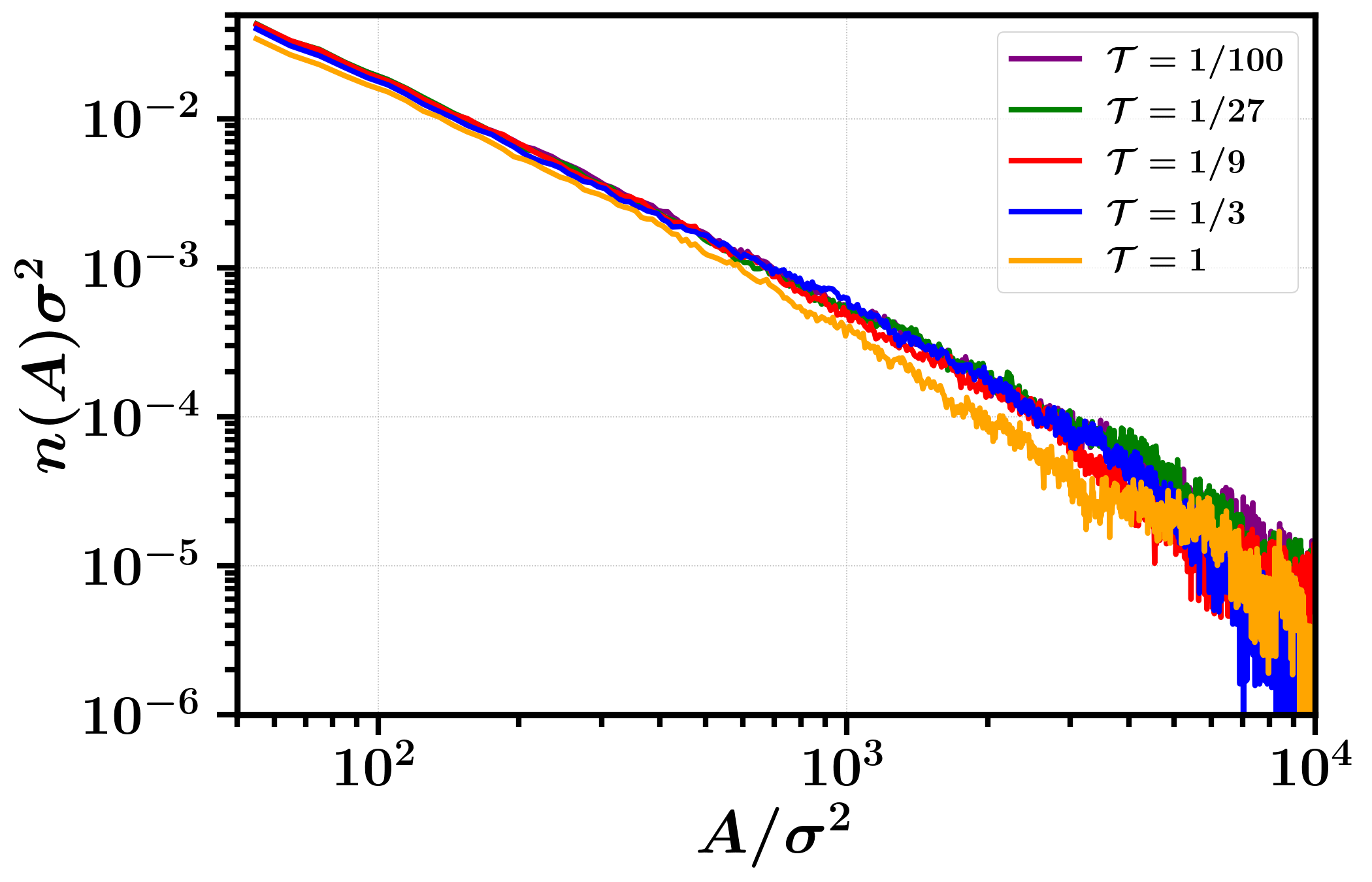}
	\caption{$n(A)$ for all monodisperse simulated values of $\mathcal{T}>0$.}
	\label{fig:numbuballthermal}
\end{figure}

\begin{figure}
	\centering
	\includegraphics[width=0.9\textwidth]{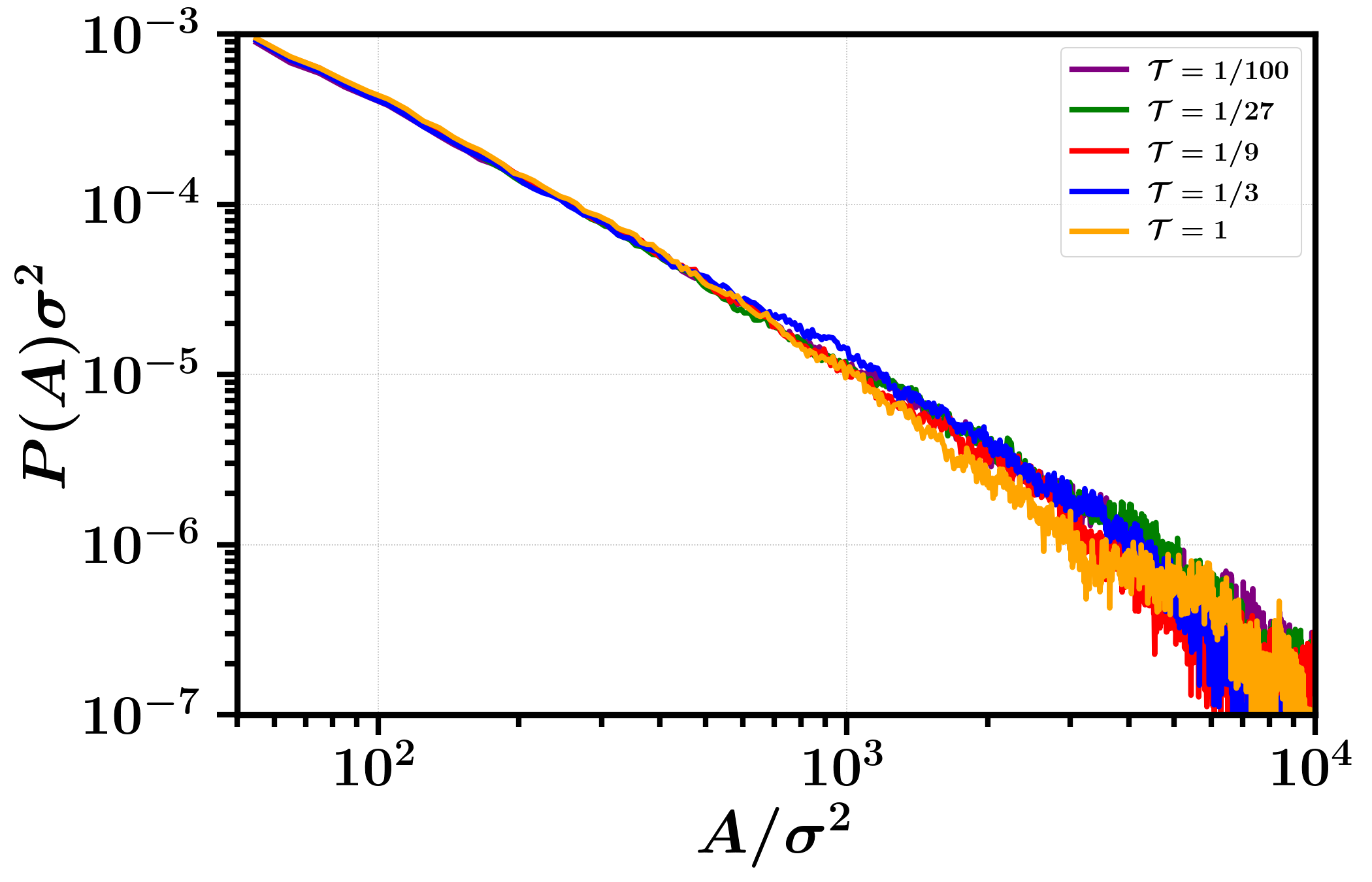}
	\caption{$P(A)$ for all monodisperse simulated values of $\mathcal{T}>0$.}
	\label{fig:pbuballthermal}
\end{figure}

\begin{figure}
	\centering
	\includegraphics[width=0.9\textwidth]{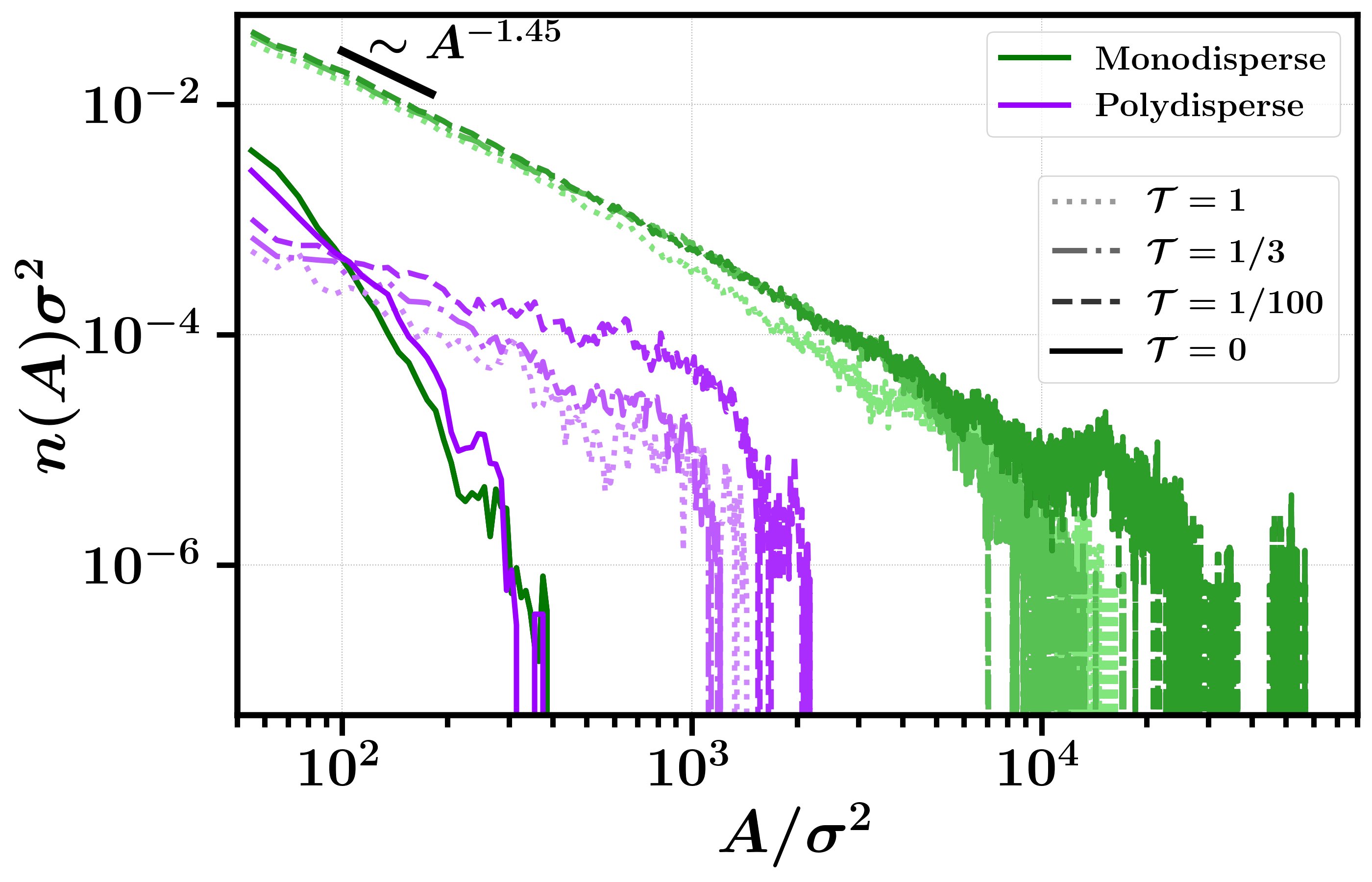}
	\caption{$n(A)$ plotted up to higher values of $A$.}
	\label{fig:numbubhighcut}
\end{figure}

We also provide $n(A)$ and $P(A)$ for all thermalized systems (not included in the main text) in Figs.~\ref{fig:numbuballthermal} and~\ref{fig:pbuballthermal}.
Additionally, we reproduce Fig. 2 of the main text up to higher values of $A$ in order to display the cutoff size for the algebraic decay in Fig.~\ref{fig:numbubhighcut}.


\subsection{\label{sec:hex}Hexatic Domains and Correlations}
The hexatic order parameter was calculated for each particle, defined as:
\begin{equation}
    \psi_{6j} = N_j^{-1}\sum_{k=1}^{N_j}\text{exp}[i6\phi_{jk}],\label{seq:hexorder}
\end{equation}
where $N_j$ is the number of nearest neighbors of particle $j$ and $\phi_{jk}$ is the angle between the vector connecting particles $j$ and $k$ with an arbitrary reference vector.
The hexatic order can be easily calculated using the order module of the Freud software package~\cite{Ramasubramani2020Freud:Data}.

Clearly the result of Eq.~\eqref{seq:hexorder} is complex, which implies a phase angle.
Following Ref.~\cite{Caporusso2020Motility-InducedSystem}, hexatic domains were defined as clusters of particles with $\psi_{6j}$ that had similar phase angle.
Specifically, the interval $[0,2\pi]$ was broken up into 15 evenly spaced bins and each particle was assigned a bin according the phase angle of its $\psi_{6j}$.
The DBSCAN algorithm outlined in the previous section was then applied while only considering particles of a particular bin with $\epsilon_{\rm cut}$ set to $1.5\sigma$ and $n_{\rm min}$ set to $6$.
The result of this calculation is clusters of particles sharing the same bin of $\psi_{6j}$ phase angle, which we identify as hexatic domains.
The area of these domains are then calculated and added to a histogram, which is then normalized by the total number of domains identified and the bin spacing $10\sigma^2$.

The above calculation was only relevant for the monodisperse simulations which displayed a grain structure.
Further, the athermal monodisperse simulation only contained a few large domains with sluggish kinetics, causing the measured distribution to be very noisy, with a few spikes corresponding to the sizes of the large grains seen in Fig. 3(a) of the main text.
We therefore only plot the domain size distribution for monodisperse thermal simulations in the main text.
We include here the same figure but with the athermal case included.

\begin{figure}
	\centering
	\includegraphics[width=0.9\textwidth]{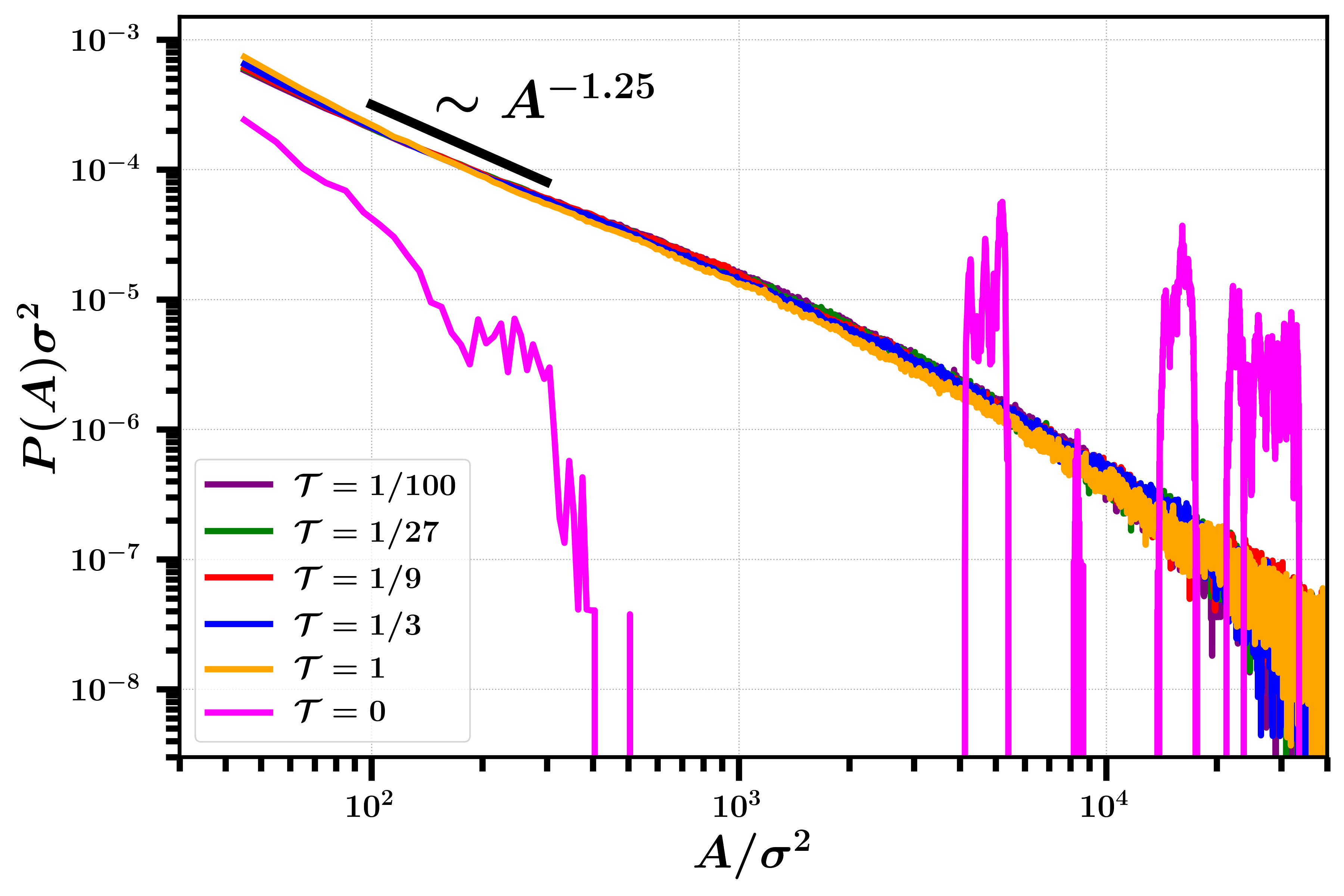}
	\caption{Grain distribution for all monodisperse simulated values, including $\mathcal{T}=0$.}
	\label{fig:graininclude}
\end{figure}

In order to confirm the disappearance of medium-to-long ranged hexatic ordering in the polydisperse case, we also calculated the spatial correlations of the hexatic order for all simulations $\langle \psi_{6}(r)\psi_{6j}^*(0) \rangle$, where $\psi_6(r)$ is the average hexatic order of all particles located within some $r+\Delta r$ of particle $j$, and the angled brackets indicate an average over all particles in the dense phase.
It should be noted that this quantity is not translationally invariant as the notation $r$ might suggest: particles closer to the interface will clearly have different hexatic order than those deep in the bulk of the dense phase.
However, in the polydisperse case where non-negligible correlations only persist for a few particle diameters, these differences minimally impact the calculation's results.

\begin{figure}
	\centering
	\includegraphics[width=0.6\textwidth]{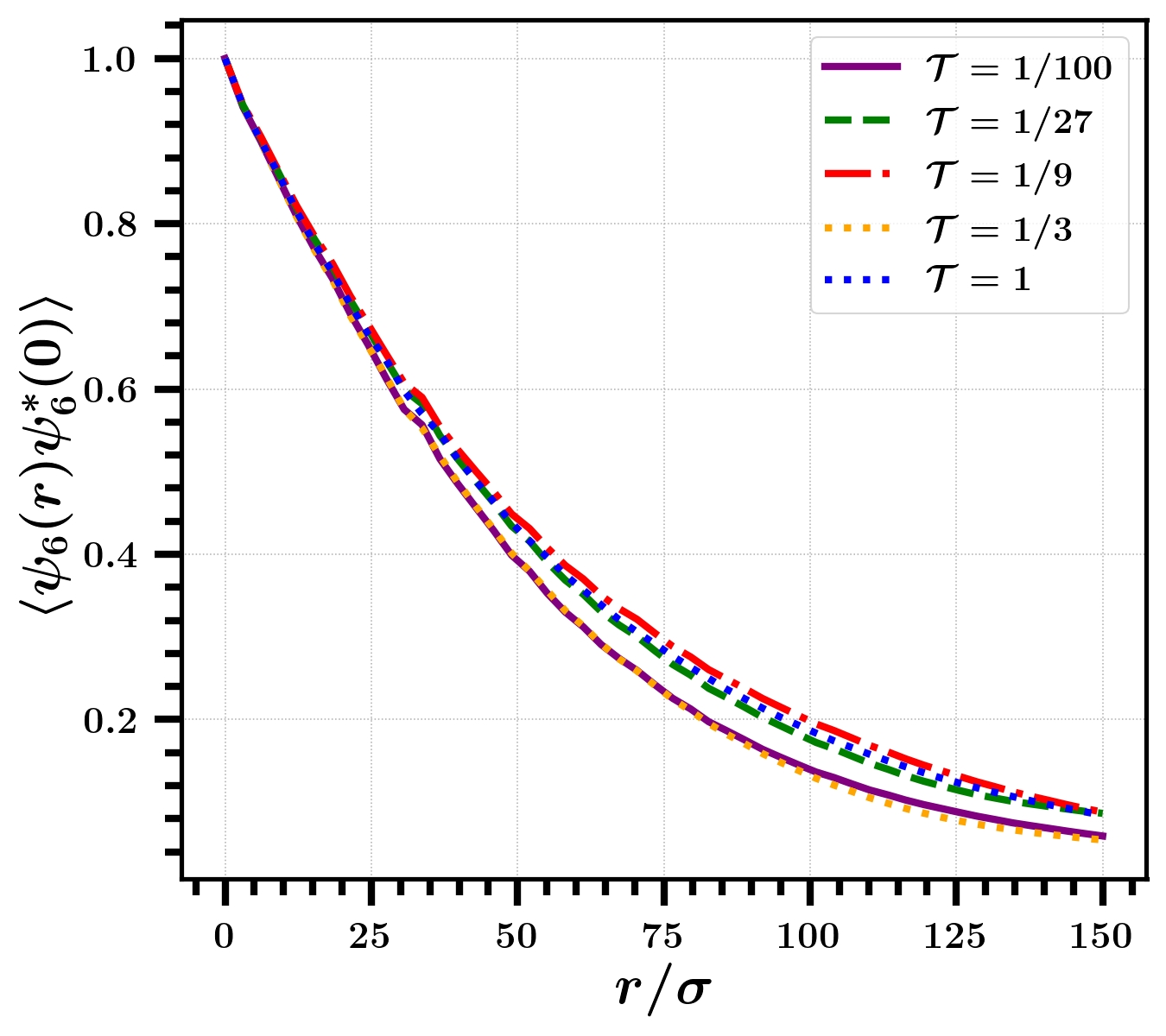}
	\caption{Hexatic spatial correlations for all monodisperse simulated values of $\mathcal{T}>0$.}
	\label{fig:hexcorrallthermal}
\end{figure}

We also provide the hexatic correlation function for all thermalized systems (not included in the main text) in Fig.~\ref{fig:hexcorrallthermal}.

\subsection{\label{sec:dyn}Dynamics}
The videos shown in Section~\ref{sec:med} point towards collective motion being a dominant transport mechanism for thermalized simulations.
In order to quantify this observation, we turn to the spatial correlation function of the velocity $Q(r)$ employed by Caprini~\textit{et al.}~\cite{Caprini2020HiddenDisks,Caprini2020SpontaneousSeparation,Caprini2021SpatialParticles}:
\begin{equation}
    Q_i(r) = 1 - 2\sum_{j}\frac{\theta_{ij}}{\mathcal{N}_r\pi},\label{seq:caprinicorr}
\end{equation}
where $\mathcal{N}_{r}$ is the number of particles in a shell of some thickness $\Delta r$ centered a distance $r$ away from the $i$th particle and $\theta_{ij}$ is the angle between the velocities of particle $i$ and particle $j$. 
Clearly, velocities are perfectly aligned $\theta_{ij}$ will be zero for all $j$, and $Q_i(r) = 1$. 
At an opposite extreme, if velocities are perfectly antialigned then $\theta_{ij} = \pi$ for all $j$, and $Q_i(r) = -1$. 
We use the DBSCAN algorithm from the previous Sections to find all particles within the dense phase, then calculate the average value of $Q(r) = (1/N_{\rm dense})\sum_{i=1}^{N_{\rm dense}}Q_i(r)$ for each particle.

\begin{figure}
	\centering
	\includegraphics[width=0.9\textwidth]{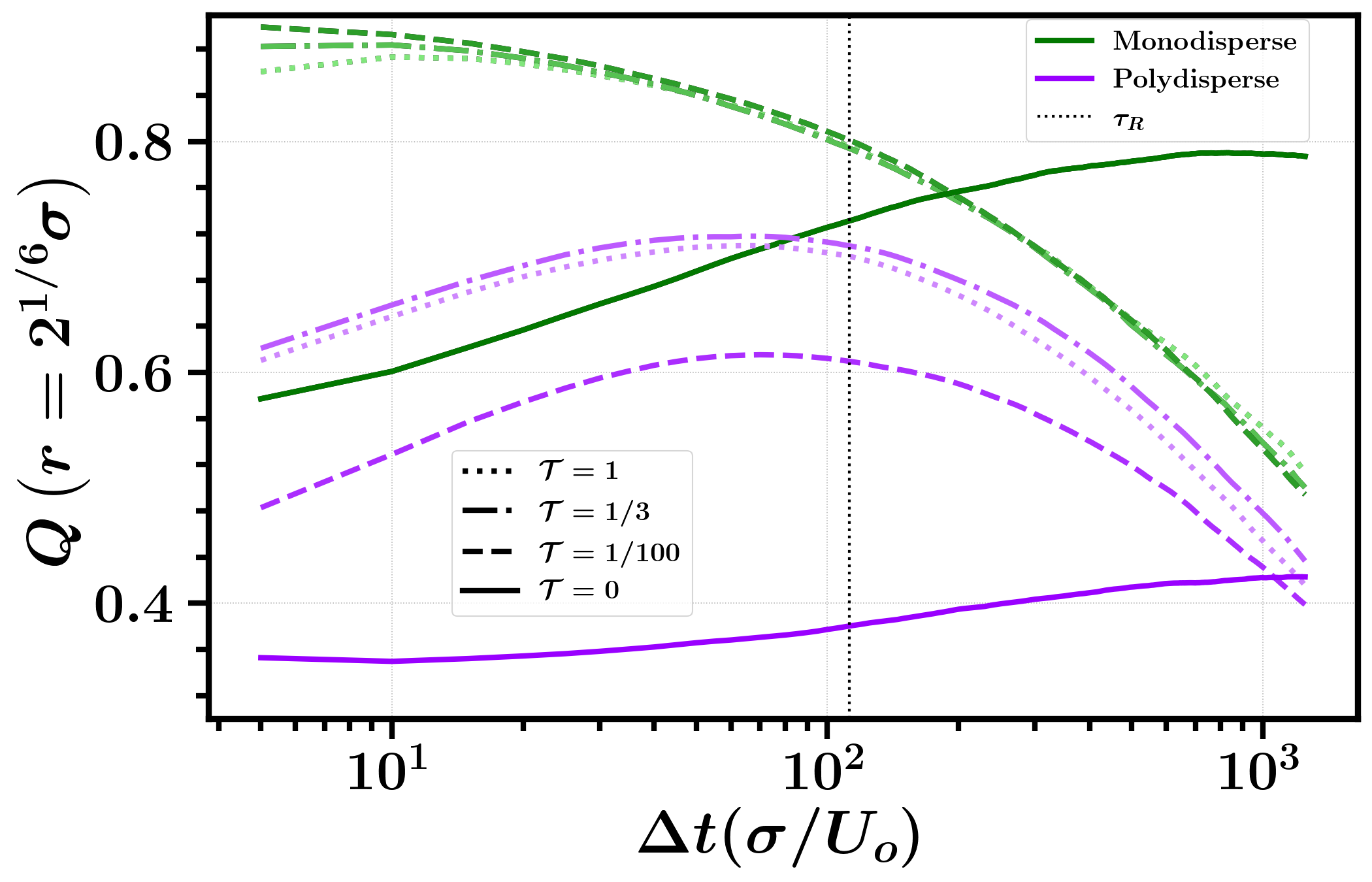}
	\caption{$Q(r)$ evaluated at the nearest neighbor separation for different selections of $\Delta t$.}
	\label{fig:qcorrdt}
\end{figure}

Note that we are analyzing Brownian data, and thus instantaneous velocities are poorly defined.
We therefore must use \textit{average} velocities $\mathbf{v}_i(t) \equiv \mathbf{r}_i(t+\Delta t) - \mathbf{r}_i(t)$ which is now a quantity dependent on the arbitrary choice of $\Delta t$.
We therefore evaluate $Q(r)$ at the distance corresponding to the first nearest neighbor shell as a function of the choice of $\Delta t$ across a wide range of time separations.
The results of this calculation are are plotted in Fig.~\ref{fig:qcorrdt}.

\begin{figure}
	\centering
	\includegraphics[width=0.6\textwidth]{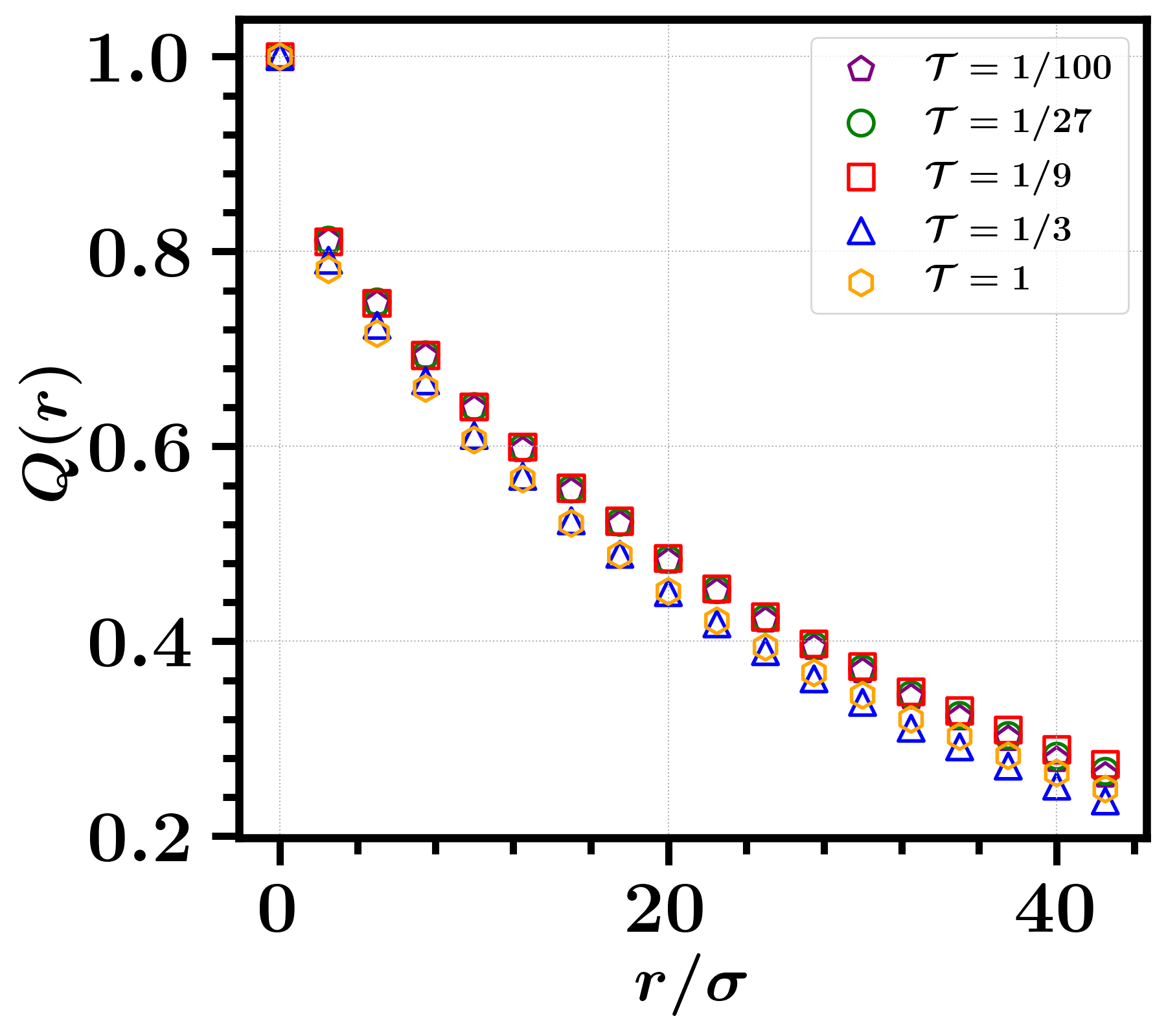}
	\caption{$Q(r)$ for all monodisperse simulated values of $\mathcal{T}>0$.}
	\label{fig:qcorrallthermal}
\end{figure}

From Fig.~\ref{fig:qcorrdt} one can determine the value of $\Delta t$ that maximizes spatial velocity correlations. 
All plots of $Q(r)$ in the main text uses the maximizing value of $\Delta t$ at each particular simulation condition.
It is interesting to note that correlations in thermal monodisperse simulations were maximized at relatively small time separations $\mathcal{O}\left(\sigma/U_o\right)$, thermal polydisperse simulations were maximized at time separations slightly less than $\mathcal{O}(\tau_R)$, and athermal simulations were maximized at time separations of multiples of $\tau_R$. 
In general, these correlations were higher for thermalized monodisperse systems, indicating that collective motion is most significant at short time scales under these conditions.
We also provide the plots of $Q(r)$ for all thermalized systems (not included in the main text) in Fig.~\ref{fig:qcorrallthermal}.

While the above provides insight on velocity correlations, they do not allow one to clearly understand the \textit{speed} at which this correlated/uncorrelated transport occurs.
We therefore calculate the mean squared displacement of particles in the dense phase.
In order to ensure that the measurement was only of dense phase particles, we employed the DBSCAN algorithm to only consider the mean squared displacement of particles that remained in the dense phase.
This systematically excludes particles which escape to the gas phase, which have typically moved a greater distance than other particles, and therefore represents an underestimate of the true dense phase MSD.
Therefore the plots in the main text which indicate subdiffusive transport in the athermal simulations may not necessarily be evidence of this.
\subsection{\label{sec:pddense}Dense Polydisperse Simulation}
All polydisperse simulations reported in the main text were done at an overall area fraction of $\phi_o = 0.49$ while all monodisperse simulations reported in the main text were done at an overall area fraction of $\phi_o = 0.64$.
In order to confirm that this discrepancy is purely quantitative and did not result in any qualitative differences, we calculated the $\mathcal{T} = 1/3$ polydisperse simulation at $\phi_o=0.64$.
The statistics of the bubbles were similar, with the area fraction of gas trapped in the bubbles $\phi_B = (1.6\pm 1.5)\times 10^{-3} $, which remains significantly smaller than that of the monodisperse thermal simulations done at similar overall densities.

$n(A)$ at this simulation is plotted along with the $n(A)$ from a simulation at the same conditions but $\phi_o=0.49$ is shown below.

\begin{figure}
	\centering
	\includegraphics[width=0.9\textwidth]{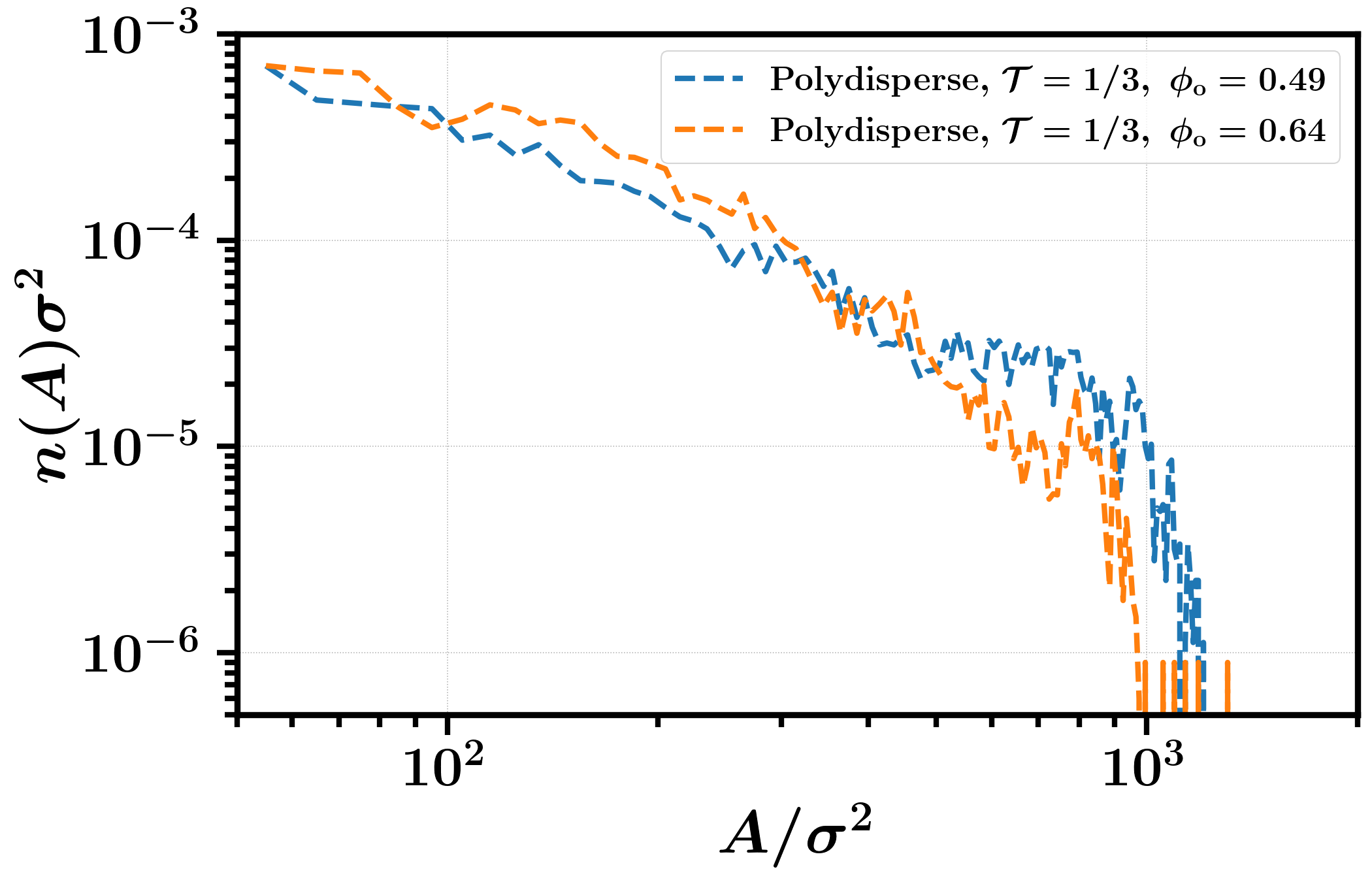}
	\caption{$n(A)$ for polydisperse $\mathcal{T} = 1/3$ simulations at $\phi_o = 0.49$ and $\phi_o = 0.64$.}
	\label{fig:bubblepdthermdense}
\end{figure}

The hexatic order spatial correlations, $Q(r)$, and the mean squared displacement were also all calculated for both the polydisperse thermal $\phi_o = 0.49$ and $\phi_o = 0.64$ simulation and are plotted below.
\begin{figure}
	\centering
	\includegraphics[width=0.9\textwidth]{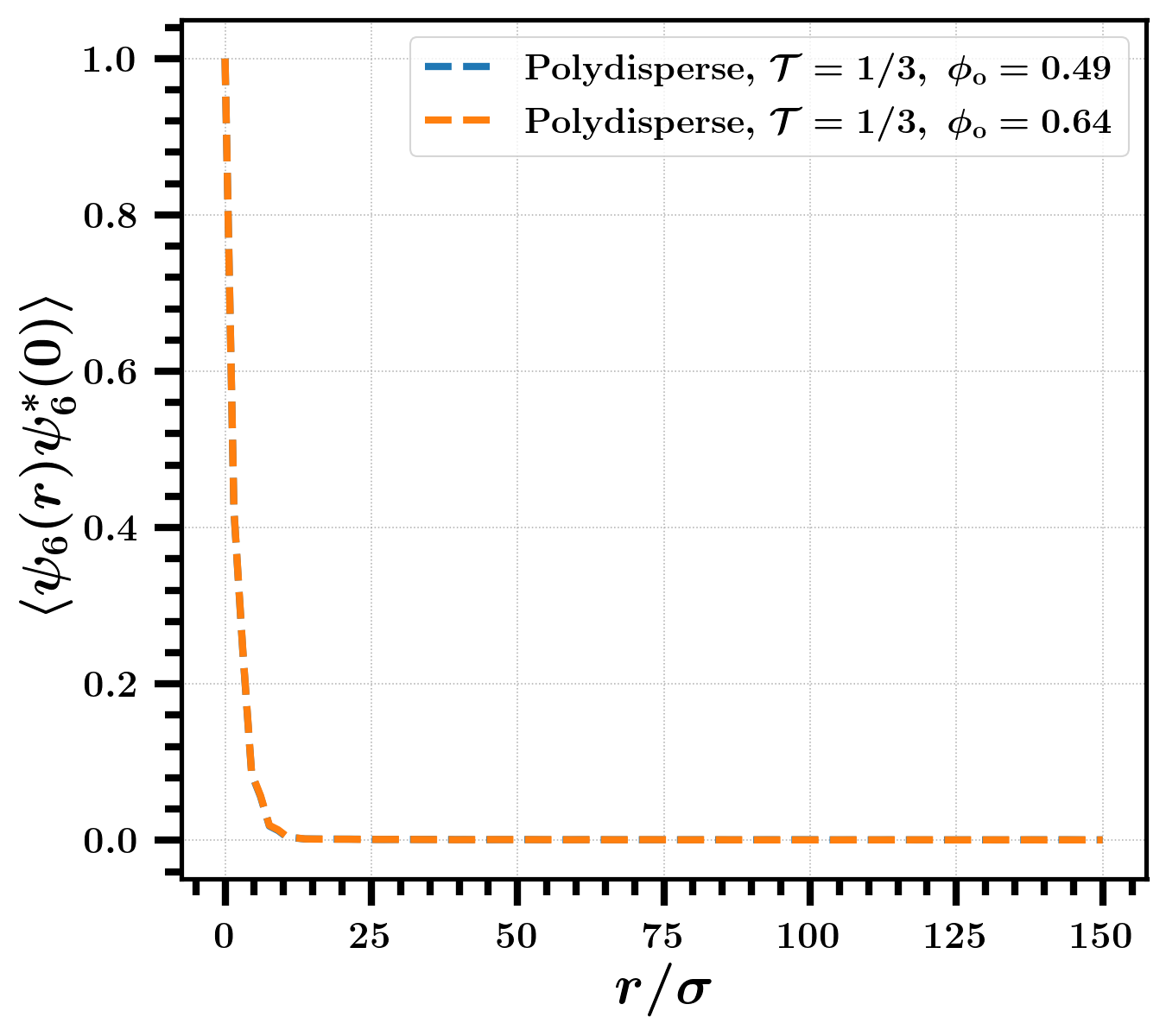}
	\caption{Spatial correlations of hexatic order for polydisperse $\mathcal{T} = 1/3$ simulations at $\phi_o = 0.49$ and $\phi_o = 0.64$.}
	\label{fig:hexpdthermdense}
\end{figure}
\begin{figure}
	\centering
	\includegraphics[width=0.9\textwidth]{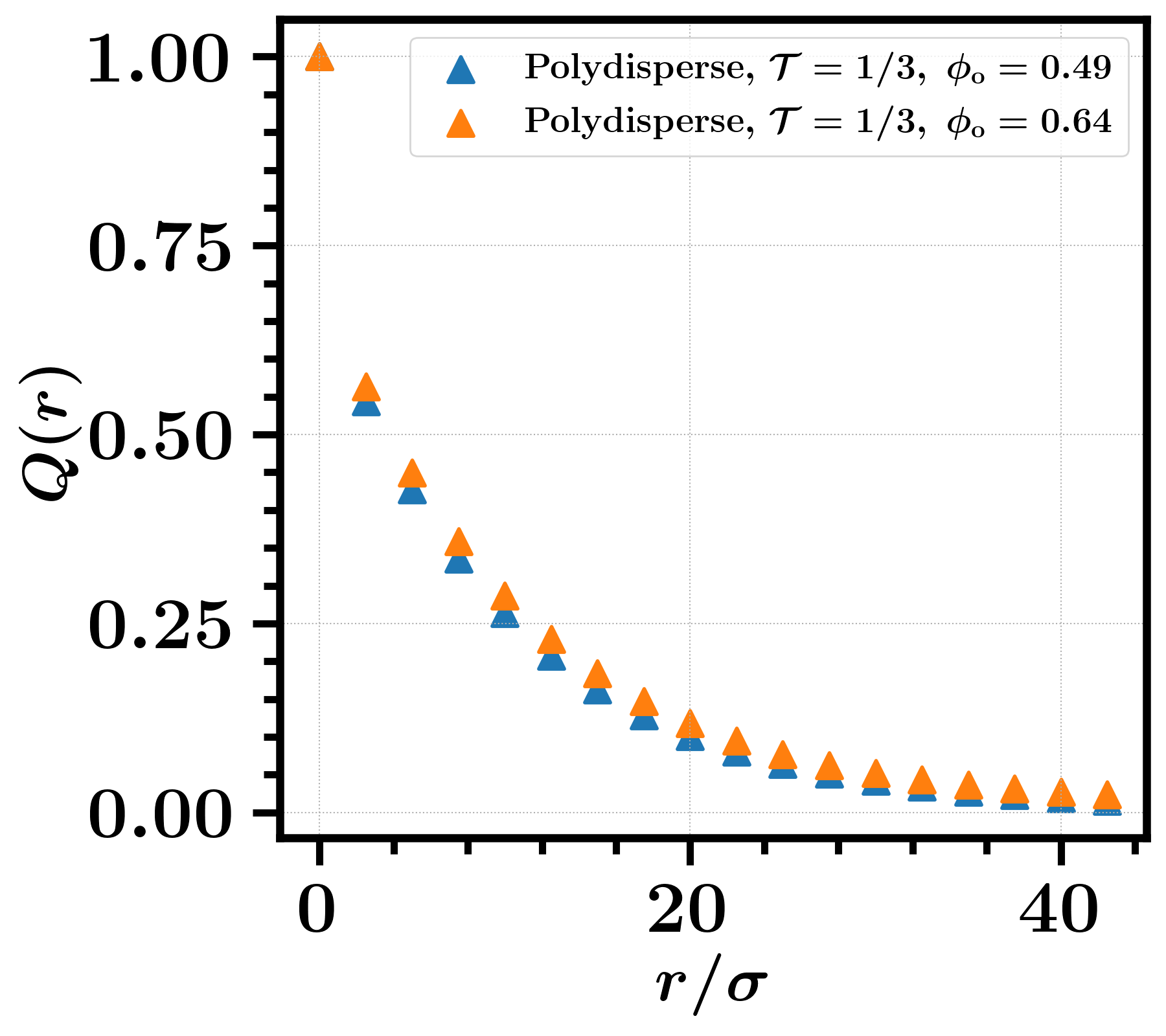}
	\caption{Spatial correlations of velocity orientation for polydisperse $\mathcal{T} = 1/3$ simulations at $\phi_o = 0.49$ and $\phi_o = 0.64$.}
	\label{fig:qcorrpdthermdense}
\end{figure}
\begin{figure}
	\centering
	\includegraphics[width=0.9\textwidth]{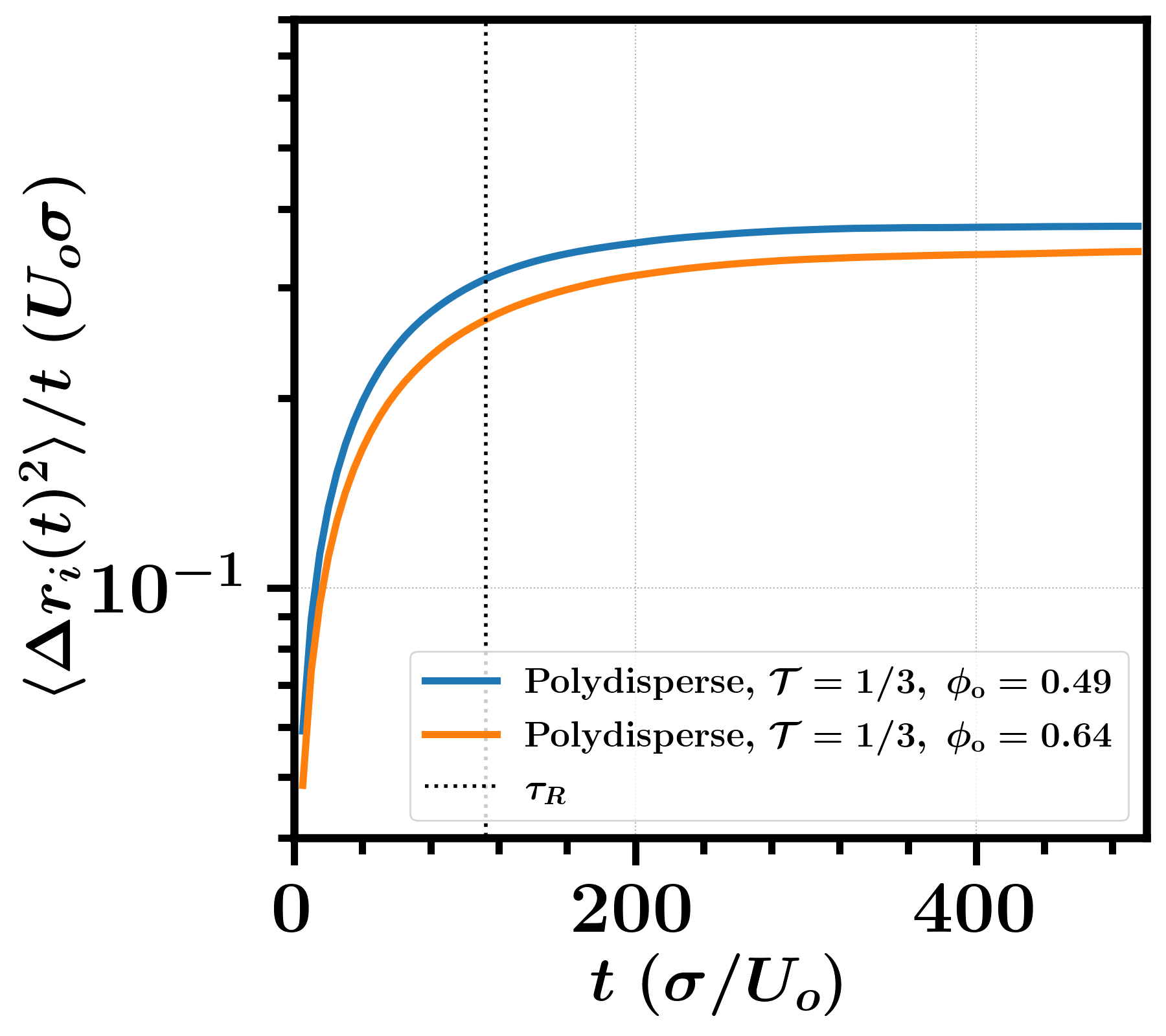}
	\caption{Mean squared displacement for polydisperse $\mathcal{T} = 1/3$ simulations at $\phi_o = 0.49$ and $\phi_o = 0.64$.}
	\label{fig:msdpdthermdense}
\end{figure} \newpage
\section{\label{sec:med}Media}
Each video listed below displays a window of selected simulations reported in the main text. 
All videos labeled ``closeup'' correspond to a total time of $2.5\times 10^{3} ~\sigma/U_o$ with frame spacing of $5~\sigma/U_o$ and all other videos correspond to a total time of $10^{4}~\sigma/U_o$ with frame spacing of $25~\sigma/U_o$.
All videos were rendered using the OVITO software package~\cite{Stukowski2010VisualizationTool} at a rate of 8 frames per second and are publicly available at the following URL:
\newline
\url{https://berkeley.box.com/s/rmjqhhvi368ae5j5fwo31b5aweeemjy9}
\newline
\begin{itemize}
    \item 
    \text{[}monodisperse\_athermal\_video.mp4\text{]}:\newline 
    Dynamics of monodisperse system with $\mathcal{T} = 0$, zoomed out such that entire system is visible.
    \item 
    \text{[}monodisperse\_athermal\_video\_closeup.mp4\text{]}:\newline 
    Dynamics of monodisperse system with $\mathcal{T}=0$, zoomed in such that motion of individual particles is visible.
    \item 
    \text{[}monodisperse\_thermal\_video.mp4\text{]}:\newline 
    Dynamics of monodisperse system with $\mathcal{T} = 1/3$, zoomed out such that entire system is visible.
    \item 
    \text{[}monodisperse\_thermal\_video\_closeup.mp4\text{]}:\newline 
    Dynamics of monodisperse system with $\mathcal{T} = 1/3$, zoomed in such that motion of individual particles is visible.
    \item 
    \text{[}polydisperse\_athermal\_video.mp4\text{]}:\newline 
    Dynamics of polydisperse system with $\mathcal{T} = 0$, zoomed out such that entire system is visible.
    \item 
    \text{[}polydisperse\_athermal\_video\_closeup.mp4\text{]}:\newline 
    Dynamics of polydisperse system with $\mathcal{T}=0$, zoomed in such that motion of individual particles is visible.
    \item 
    \text{[}polydisperse\_thermal\_video.mp4\text{]}:\newline
    Dynamics of polydisperse system with $\mathcal{T}=1/3$, zoomed out such that entire system is visible.
    \item 
    \text{[}polydisperse\_thermal\_video\_closeup.mp4\text{]}:\newline 
    Dynamics of polydisperse system with $\mathcal{T}=1/3$, zoomed in such that motion of individual particles is visible.
\end{itemize}